\documentclass[ALICE,manyauthors]{cernphprep}
\usepackage[comma,square,numbers,sort&compress]{natbib}
\usepackage{hyperref}
\usepackage{lineno}
\usepackage{xspace}
\usepackage{xcolor}
\usepackage{soul} 
\usepackage{mathtools}
\usepackage{multirow}
\usepackage{amsmath}
\usepackage{graphicx}
\usepackage{comment}
\usepackage[utf8]{inputenc}
\usepackage{csquotes}
\usepackage[T1]{fontenc} 
\usepackage{orcidlink}

\begin{document}
%

\newcommand{\pp}           {pp\xspace}
\newcommand{\ppbar}        {\mbox{$\mathrm {p\overline{p}}$}\xspace}
\newcommand{\XeXe}         {\mbox{Xe--Xe}\xspace}
\newcommand{\PbPb}         {\mbox{Pb--Pb}\xspace}
\newcommand{\pA}           {\mbox{pA}\xspace}
\newcommand{\pPb}          {\mbox{p--Pb}\xspace}
\newcommand{\AuAu}         {\mbox{Au--Au}\xspace}
\newcommand{\dAu}          {\mbox{d--Au}\xspace}

\newcommand{\s}            {\ensuremath{\sqrt{s}}\xspace}
\newcommand{\snn}          {\ensuremath{\sqrt{s_{\mathrm{NN}}}}\xspace}
\newcommand{\pt}           {\ensuremath{p_{\rm T}}\xspace}
\newcommand{\meanpt}       {$\langle p_{\mathrm{T}}\rangle$\xspace}
\newcommand{\ycms}         {\ensuremath{y_{\rm CMS}}\xspace}
\newcommand{\ylab}         {\ensuremath{y_{\rm lab}}\xspace}
\newcommand{\etarange}[1]  {\mbox{$\left | \eta \right |~<~#1$}}
\newcommand{\yrange}[1]    {\mbox{$\left | y \right |~<~#1$}}
\newcommand{\dndy}         {\ensuremath{\mathrm{d}N/\mathrm{d}y}\xspace}
\newcommand{\dndeta}       {$\langle \mathrm{d}N_\mathrm{ch}/\mathrm{d}\eta \rangle_{|\eta|<0.5}$\xspace}
\newcommand{\dndetaqube}   {$\langle \mathrm{d}N_\mathrm{ch}/\mathrm{d}\eta \rangle^{1/3}_{|\eta|<0.5}$\xspace}
\newcommand{\avdndeta}     {\ensuremath{\langle\dndeta\rangle}\xspace}
\newcommand{\dNdy}         {\ensuremath{\mathrm{d}N_\mathrm{ch}/\mathrm{d}y}\xspace}
\newcommand{\Npart}        {\ensuremath{N_\mathrm{part}}\xspace}
\newcommand{\Ncoll}        {\ensuremath{N_\mathrm{coll}}\xspace}
\newcommand{\dEdx}         {\ensuremath{\textrm{d}E/\textrm{d}x}\xspace}
\newcommand{\RpPb}         {\ensuremath{R_{\rm pPb}}\xspace}

\newcommand{\nineH}        {$\sqrt{s}~=~0.9$~Te\kern-.1emV\xspace}
\newcommand{\seven}        {$\sqrt{s}~=~7$~Te\kern-.1emV\xspace}
\newcommand{\twoH}         {$\sqrt{s}~=~0.2$~Te\kern-.1emV\xspace}
\newcommand{\twosevensix}  {$\sqrt{s}~=~2.76$~Te\kern-.1emV\xspace}
\newcommand{\five}         {$\sqrt{s}~=~5.02$~Te\kern-.1emV\xspace}
\newcommand{\twosevensixnn}{$\sqrt{s_{\mathrm{NN}}}~=~2.76$~Te\kern-.1emV\xspace}
\newcommand{\fivenn}       {$\sqrt{s_{\mathrm{NN}}}~=~5.02$~Te\kern-.1emV\xspace}
\newcommand{\LT}           {L{\'e}vy-Tsallis\xspace}
\newcommand{\GeVc}         {Ge\kern-.1emV/$c$\xspace}
\newcommand{\MeVc}         {Me\kern-.1emV/$c$\xspace}
\newcommand{\TeV}          {Te\kern-.1emV\xspace}
\newcommand{\GeV}          {Ge\kern-.1emV\xspace}
\newcommand{\MeV}          {Me\kern-.1emV\xspace}
\newcommand{\GeVmass}      {Ge\kern-.2emV/$c^2$\xspace}
\newcommand{\MeVmass}      {Me\kern-.2emV/$c^2$\xspace}
\newcommand{\lumi}         {\ensuremath{\mathcal{L}}\xspace}
\newcommand{\rap}         {$|\it{y}|$\xspace}

\newcommand{\ITS}          {\rm{ITS}\xspace}
\newcommand{\TOF}          {\rm{TOF}\xspace}
\newcommand{\ZDC}          {\rm{ZDC}\xspace}
\newcommand{\ZDCs}         {\rm{ZDCs}\xspace}
\newcommand{\ZNA}          {\rm{ZNA}\xspace}
\newcommand{\ZNC}          {\rm{ZNC}\xspace}
\newcommand{\SPD}          {\rm{SPD}\xspace}
\newcommand{\SDD}          {\rm{SDD}\xspace}
\newcommand{\SSD}          {\rm{SSD}\xspace}
\newcommand{\TPC}          {\rm{TPC}\xspace}
\newcommand{\TRD}          {\rm{TRD}\xspace}
\newcommand{\VZERO}        {\rm{V0}\xspace}
\newcommand{\VZEROA}       {\rm{V0A}\xspace}
\newcommand{\VZEROC}       {\rm{V0C}\xspace}
\newcommand{\Vdecay} 	   {\ensuremath{V^{0}}\xspace}

\newcommand{\ee}           {\ensuremath{e^{+}e^{-}}} 
\newcommand{\pip}          {\ensuremath{\pi^{+}}\xspace}
\newcommand{\pim}          {\ensuremath{\pi^{-}}\xspace}
\newcommand{\kap}          {\ensuremath{\rm{K}^{+}}\xspace}
\newcommand{\kam}          {\ensuremath{\rm{K}^{-}}\xspace}
\newcommand{\pbar}         {\ensuremath{\rm\overline{p}}\xspace}
\newcommand{\kzero}        {\ensuremath{{\rm K}^{0}_{\rm{S}}}\xspace}
\newcommand{\lmb}          {\ensuremath{\Lambda}\xspace}
\newcommand{\almb}         {\ensuremath{\overline{\Lambda}}\xspace}
\newcommand{\Om}           {\ensuremath{\Omega^-}\xspace}
\newcommand{\Mo}           {\ensuremath{\overline{\Omega}^+}\xspace}
\newcommand{\X}            {\ensuremath{\Xi^-}\xspace}
\newcommand{\Ix}           {\ensuremath{\overline{\Xi}^+}\xspace}
\newcommand{\Xis}          {\ensuremath{\Xi^{\pm}}\xspace}
\newcommand{\Oms}          {\ensuremath{\Omega^{\pm}}\xspace}
\newcommand{\degree}       {\ensuremath{^{\rm o}}\xspace}
\newcommand{\Kstar}         {\ensuremath{\mathrm{K^{*0}}}\xspace}
\newcommand{\Kstarpdg}         {\ensuremath{\mathrm{K^{*}(892)^{0}}}\xspace}
\newcommand{\Rhopdg}         {\ensuremath{\mathrm{\rho(770)^{0}}}\xspace}
\newcommand{\Lambdapdg}         {\ensuremath{\mathrm{\Lambda(1520)}}\xspace}
\newcommand{\Phipdg}         {\ensuremath{\mathrm{\phi(1020)}}\xspace}
\newcommand{\Phimeson}         {\ensuremath{\mathrm{\phi}}\xspace}
\newcommand{\NKS}          {K(892)$^{*0}$\xspace}
\newcommand{\NKSshort}          {K$^{*0}$\xspace}
\newcommand{\ENxe}        {$\sqrt{\it{s}_{\mathrm{NN}}} =$ 5.44 TeV\xspace}
\newcommand{\ENpp}        {$\sqrt{\it{s}} =$ 5.02 TeV\xspace}
\newcommand{\pion}          {$\mathrm{\pi}$\xspace}
\newcommand{\ENpbpb}        {$\sqrt{\it{s}_{\mathrm{NN}}} =$ 5.02 TeV\xspace}
\newcommand{\MPT}        {$\langle p_{\mathrm{T}} \rangle$\xspace}

\begin{titlepage}
\PHyear{2023}       
\PHnumber{175}      
\PHdate{22 August}  

\title{System size dependence of the hadronic rescattering effect at energies available at the CERN Large Hadron Collider}
\ShortTitle{Probing hadronic rescattering effect across different systems at LHC energies}   

\Collaboration{ALICE Collaboration\thanks{See Appendix~\ref{app:collab} for the list of collaboration members}}
\ShortAuthor{ALICE Collaboration} 

\begin{abstract}

The first measurements of \Kstarpdg resonance production as a function of charged-particle multiplicity in \XeXe collisions at \snn $=$ 5.44 TeV and pp collisions at \s $=$ 5.02 TeV using the ALICE detector are presented. The resonance is reconstructed at midrapidity (\rap $<$ 0.5) using the hadronic decay channel \Kstar $\rightarrow \mathrm{K^{\pm} \pi^{\mp}}$. Measurements of transverse-momentum integrated yield, mean transverse-momentum, nuclear modification factor of \Kstar, and yield ratios of resonance to stable hadron (\Kstar/K) are compared across different collision systems (pp, \pPb, \XeXe, and \PbPb) at similar collision energies to investigate how the production of \Kstar resonances depends on the size of the system formed in these collisions. The hadronic rescattering effect is found to be independent of the size of colliding systems and mainly driven by the produced charged-particle multiplicity, which is a proxy of the volume of produced matter at the chemical freeze-out. In addition, the production yields of \Kstar in \XeXe collisions are utilized to constrain the dependence of the kinetic freeze-out temperature on the system size using the hadron resonance gas in partial chemical equilibrium (HRG-PCE) model. 
	
\end{abstract}
\end{titlepage}

\setcounter{page}{2} 


\section{Introduction} 
\label{Section:Introduction}  

Production of hadrons consisting of light-flavored quarks (u, d and s) have been extensively studied in heavy-ion collisions as well as in small collision systems like pp and p--Pb~\cite{ALICE:2013wgn, ALICE:2013mez, ALICE:2015mpp, ALICE:2011gmo, ALICE:2016sak, ALICE:2021ptz, ALICE:2018pal, ALICE:2019avo} at CERN Large Hadron Collider (LHC) energies to investigate the bulk properties of strongly interacting quantum chromodynamics (QCD) matter of deconfined quarks and gluons, known as the quark--gluon plasma (QGP)~\cite{STAR:2000ekf, STAR:2005gfr, STAR:2003pjh, STAR:2002svs, STAR:2003wqp, PHENIX:2001hpc, PHENIX:2004vcz, BRAHMS:2004adc, PHOBOS:2004zne, ALICE:2022wpn, ALICE:2010suc, ALICE:2010yje, ALICE:2011ab, Heinz:2008tv, Niida:2021wut}. The produced QGP is modelled by hydrodynamical equations~\cite{Blaizot:1987cc, Teaney:2009qa}. The system while evolving cools down, and after a certain time, hadronization takes place~\cite{Greco:2003xt, Greco:2003mm,Fries:2003vb,Fries:2003kq, Andronic:2017pug, HotQCD:2018pds}. As the temperature of the system dials down further, it first reaches a space--time surface called chemical freeze-out surface~\cite{Teaney:2002aj} where the hadronic abundances get fixed, and then a kinetic freeze-out ($T_{\mathrm{kin}}$) surface where the hadrons momenta get frozen~\cite{Manninen:2008mg, Heinz:2007in}. After the kinetic freeze-out surface, particles stream freely to the detectors. 
In these collisions, several kinds of light and heavy flavor hadrons and resonances with different flavors of valence quark content, mass, and lifetime are produced. Each of these hadrons and resonances possesses unique characteristic features that can be exploited to study the properties of the medium. Hadron yields are used as an experimental input in the thermal model~\cite{Karthein:2022fcp, Keranen:2001un, Vovchenko:2019kes, Cleymans:2010qya, DasGupta:1981xx} to extract the chemical freeze-out temperature, baryon chemical potential, and volume of the produced matter. The transverse-momentum (\pt) spectra of hadrons are fitted with a hydrodynamics-based model, such as the blast-wave model~\cite{Schnedermann:1993ws}, to obtain the kinetic freeze-out temperature~\cite{Heinz:2007in, ALICE:2019hno} and collective radial expansion velocity~\cite{ALICE:2019hno} of the medium. The phase between the chemical and kinetic freeze-out surface is termed as the hadronic phase~\cite{Steinheimer:2017vju}. Properties of the hadronic phase can be probed by studying short-lived resonance particles which decay via the strong interaction. 
Short-lived resonances have a lifetime comparable to that of the hadronic phase and, therefore, their decay products get engaged in regeneration~\cite{Vogel:2005qr, STAR:2006vhb} and rescattering~\cite{ALICE:2019xyr, STAR:2004bgh} processes. These processes depend on the hadronic cross section~\cite{Kopeliovich:2014wxa, Aston:1987ir, Protopopescu:1973sh} of the decay products of the resonance inside the hadronic medium, the lifetime of the resonance particle, density of the hadron gas, and the hadronic phase lifetime. The presence of these final-state hadronic interactions leads to the modification of experimentally measured yields of resonance particles~\cite{Torrieri:2001ue, Markert:2005jv}.

To probe the final-state hadronic interactions, extensive study for the production of light flavor resonances with different lifetimes ($\tau$), e.g.\ \Kstar ($\tau\eqsim$ 4 fm/$c$) and \Phipdg ($\tau\eqsim$ 40 fm/$c$) has been carried out previously in various collision systems~\cite{ALICE:2021ptz, ALICE:2014jbq, ALICE:2019etb, ALICE:2021uyz, ALICE:2022uac, NA49:2008goy, NA49:2000jee, NA49:2011bfu, PHENIX:2014kia, PHENIX:2022rvg, PHENIX:2022hku, PHENIX:2010bqp, PHENIX:2004spo, STAR:2008inc, STAR:2008bgi, STAR:2002npn, STAR:2004bgh, STAR:2010avo}. The \pt-integrated yield of \Kstar relative to kaons is found to be suppressed in central heavy-ion collisions compared to pp, peripheral heavy-ion collisions, and to thermal model predictions, whereas no such suppression is observed for the \Phimeson meson. The observed reduction in measurable yield suggests that the rescattering of \Kstar decay products in the hadronic phase dominates over regeneration, leading to the suppression of measurable yield. The suppression of \Kstar meson yields due to the rescattering is dominant at low \pt($<$ 3 GeV/$c$) from the study of \pt-differential \Kstar/K yield ratio~\cite{ALICE:2019xyr}. Furthermore, at high \pt, the phenomenon of energy loss by energetic partons traversing the dense medium formed in high-energy heavy-ion collisions affects the production yield of \Kstar and \Phimeson resonances~\cite{ALICE:2017ban, ALICE:2021ptz} compared to the pp collisions. The energy loss process depends on the lifetime of the dense matter, initial medium density, the path length traversed by the parton, and the parton flavor. The modification in the yield of high-\pt particles is quantified using the nuclear modification factor ($R_{\mathrm{AA}}$)~\cite{PHENIX:2001hpc} defined as\\
\begin{equation}
	R_{\mathrm{AA}} = \frac{1}{\langle T_{\mathrm{AA}} \rangle}\frac{\mathrm{d}^{2}N^{\mathrm{AA}}/(\mathrm{d}y\mathrm{d}p_{\mathrm{T}})}{\mathrm{d}^{2}\sigma^{\mathrm{pp}}/(\mathrm{d}y\mathrm{d}p_{\mathrm{T}})},
\end{equation}
where $\mathrm{d}^{2}N^{\mathrm{AA}}/(\mathrm{d}y\mathrm{d}p_{\mathrm{T}})$ is the yield of the particle in heavy-ion collisions and $\sigma^{\mathrm{pp}}$ is its production cross section in pp collisions. The average nuclear overlap function is  denoted by  $\langle T_{\mathrm{AA}} \rangle$ and can be estimated as $\langle T_{\mathrm{AA}} \rangle = \langle N_{\mathrm{coll}} \rangle / \sigma_{\mathrm{inel}}$, where $\langle N_{\mathrm{coll}} \rangle$ is the average number of binary nucleon--nucleon collisions obtained from Monte Carlo Glauber simulations~\cite{Loizides:2014vua} and $\sigma_{\mathrm{inel}}$ is the inelastic pp cross section~\cite{Loizides:2017ack}. The $R_{\mathrm{AA}}$ measurements for \Kstar and \Phimeson in Pb--Pb collisions at \snn $=$ 2.76 and 5.02 TeV show that at high \pt ($>$ 6 GeV/$c$) energy loss for $\mathrm{\pi}$, K, p, \Kstar and \Phimeson are consistent within uncertainties. This observation suggests that the partonic energy loss in the QGP medium is independent of the flavor of light quarks (u, d, s)~\cite{ALICE:2019hno, ALICE:2021ptz, ALICE:2017ban}.

Recent measurements of light flavor hadron production in high-multiplicity pp and p--Pb collisions show some characteristics~\cite{ALICE:2016fzo, ALICE:2021nir, ALICE:2019zfl, PHENIX:2018lia, CMS:2016fnw} which have so far been solely attributed to the medium created in heavy-ion collisions. The systems created in pp, p--Pb, and heavy-ion collisions, can be classified based on the final state average charged-particle pseudorapidity density measured at midrapidity (\dndeta). In small collision systems, multiplicities range from a few to a few tens of charged particles per unit of pseudorapidity. In contrast, in Pb--Pb collisions, multiplicities of a few thousand charged-particles per unit of rapidity can be produced. Recent studies by the ALICE Collaboration at LHC energies show a smooth evolution of the yield or abundance of different hadron species as a function of \dndeta across different collision systems and energies~\cite{ALICE:2018pal,ALICE:2019avo}. In contrast, the mean transverse-momentum (\meanpt), which depends on the radial flow,
follows a different trend across various colliding systems, rising faster in small collision systems (pp, p--Pb) compared to heavy-ion (Pb--Pb) collisions~\cite{ALICE:2019avo}. One of the primary motivations for studying resonances like \Kstar and \Phimeson in high-multiplicity pp and p--Pb collisions is to search for the presence of a hadronic phase with a non-zero lifetime in a small collision system. A hint of suppression of \Kstar meson production in high-multiplicity pp and p--Pb collisions was previously reported by the ALICE Collaboration~\cite{ALICE:2019etb}. In fact, the suppression of \Kstar/K yield ratio evolves smoothly as a function of \dndeta from low-multiplicity pp collisions to central Pb--Pb collisions across different collision energies.

The measurement of \Kstar production yield in the collisions of medium-sized nuclei such as Xe--Xe provides a test for validating the picture of the smooth evolution of hadronic rescattering across different colliding systems by bridging the gap between p--Pb and Pb--Pb multiplicities. Using the data sets of pp, p--Pb, Xe--Xe, and Pb--Pb collisions, collected by the ALICE Collaboration at center-of-mass energies per nucleon pair (\snn) of about 5 TeV, a systematic study of system-size dependence of hadronic rescattering is possible. In this article, the first measurements of \Kstar meson production at midrapidity ($|y|<$ 0.5) as a function of \dndeta in pp collisions at \s $=$ 5.02 TeV and in \XeXe collisions at \snn $=$ 5.44 TeV are presented. The measured \Kstar yield and \Kstar/K yield ratio in these collisions are compared with the results obtained from p--Pb and Pb--Pb collisions to understand the system-size dependency of \Kstar production and the hadronic rescattering effect. The yield ratio \Kstar/K is used to constrain the hadronic phase lifetime across different collision systems. Furthermore, the measured \Kstar yields in \XeXe and \PbPb collisions are used as an experimental input in a partial chemical equilibrium (PCE) based thermal model HRG-PCE~\cite{Motornenko:2019jha} to constrain the kinetic freeze-out temperature. This is a novel procedure to extract $T_{\mathrm{kin}}$ that is independent of assumptions about the flow velocity profile and the freeze-out hypersurface~\cite{Motornenko:2019jha}. In addition, the mean values of transverse-momentum (\meanpt) of \Kstar in different collision systems are also compared to understand the evolution of radial flow from small collision systems to heavy-ion collisions. Moreover, the $R_{\mathrm{AA}}$ of \Kstar at similar charged-particle multiplicity in Pb$-$Pb and Xe$-$Xe collisions are compared to shed light on the system-size dependence of parton energy loss.

The organization of the article is as follows. The ALICE experimental setup, data analysis technique, and sources of systematic uncertainties are described in Secs.~\ref{EA},~\ref{DA} and~\ref{Section:Systematics}, respectively. Results are shown in Sec.~\ref{Res}, and the article is finally summarized in Sec.~\ref{Section:Summary}. Since the production of particles and antiparticles are in equal amounts at midrapidity at LHC energies~\cite{ALICE:2019hno}, the results for \Kstar and $\overline{\mathrm{K}^{*0}}$ are averaged and denoted as \Kstar throughout the article unless stated otherwise.

\section{Experimental apparatus, event and track selection} 
\label{EA}

The production yield of \Kstar meson is measured in Xe$-$Xe and pp collisions at \ENxe and \ENpp, respectively, using the data collected by the ALICE detector at the LHC. The \XeXe collision events were collected in the year 2017 with a magnetic field strength B $=$ 0.2 T, whereas the pp collision events were collected with B $=$ 0.5 T in the year 2015. A full description of the ALICE detector and its performance can be found in~\cite{ALICE:2014sbx, ALICE:2008ngc}. Analyzed events are selected using a minimum-bias trigger that requires at least one hit in both forward scintillator detectors V0A (2.8$ < \eta <$ 5.1) and V0C ($-$3.7$ < \eta < -$1.7)~\cite{ALICE:2013axi}. 
Pileup removal involves analyzing hits in the SPD detector, correlating cluster numbers in the ITS and TPC detectors, identifying multiple vertices with the SPD detector, and utilizing the correlation between the SPD and V0M detectors. Beam-induced background and pileup events are eliminated through an offline event selection process, as described in Refs.~\cite{ALICE:2014sbx, ALICE:2019avo} for pp and~\cite{ALICE:2018hza, ALICE:2019hno} for \XeXe collisions. The results for pp collisions presented in this paper are based on the \enquote{INEL $>$ 0} event class, which is a subset of inelastic collisions where at least one charged particle is emitted in the pseudorapidity interval $|\eta|<$ 1~\cite{ALICE:2015qqj}. In addition, selected events must have one primary collision vertex which is reconstructed using the two innermost layers of the Inner Tracking System (ITS)~\cite{ALICE:2013nwm} and is located within $\pm$10 cm along the beam axis from the nominal center of the ALICE detector. Measurements for \Kstar production yields are carried out using 1.44 $\times$ 10$^{6}$ and 100 $\times$ 10$^{6}$ minimum-bias \XeXe and pp collision events. The selected events are categorized into distinct classes based on their centrality in heavy-ion collisions (\XeXe) or multiplicity in proton--proton (pp) collisions. These event classes are defined using percentiles of the hadronic cross section. The classification of event classes is accomplished by analyzing the signal deposited in both V0 detectors, referred to as the \enquote{V0M amplitude}, which is proportional to the charged-particle multiplicity. Various measured observables, such as the transverse momentum (\pt) spectrum, transverse-momentum-integrated yield (\dndy), mean transverse momentum (\meanpt), yield ratios of resonances to stable particles, kinetic freeze-out temperature ($T_{\mathrm{kin}}$), and nuclear modification factor ($R_{\mathrm{AA}}$), are presented for different multiplicity (or centrality for heavy-ion collisions) classes as a function of pseudorapidity density (\dndeta)~\cite{ALICE:2018cpu, ALICE:2019dfi}.

In \XeXe collisions, the measurements are conducted in four different centrality classes: 0--30\%, 30--50\%, 50--70\%, and 70--90\%. The centrality classes of 0--30\% and 70--90\% represent central and peripheral collisions, respectively. On the other hand, in pp collisions, the measurements are performed in nine different multiplicity classes, as listed in Table~\ref{tab:mult}, with class I having the highest multiplicity and class IX having the lowest multiplicity.
\begin{table}[ht]
	\centering
	\caption{Analyzed multiplicity classes in pp collisions at \s $=$ 5.02 TeV}
	
	\begin{tabular}{|c|c|c|c|c|c|c|c|c|c|}	
		\hline
		\textbf{V0M ($\%$)} & 0--1 & 1--5 & 5--10 & 10--20 & 20--30 & 30--40 & 40--50 & 50--70 & 70--100\\ 
		\hline
		\textbf{Multiplicity classes}& I & II & III & IV & V & VI & VII & VIII & IX\\
		\hline   
		
	\end{tabular}
	\label{tab:mult}	
\end{table}

Charged tracks in a selected event are reconstructed using the ITS~\cite{ALICE:2013nwm} and Time Projection Chamber (TPC)~\cite{Alme:2010ke} detectors, which are located within a solenoid that provides a homogeneous magnetic field. In order to ensure good track quality, a set of track selection criteria are used, as done in previous works~\cite{ALICE:2019etb, ALICE:2021lsv}. Charged tracks coming from the primary collision vertex are selected with minimum \pt of 0.15 GeV/$c$ and $|\eta|<$ 0.8. Selected tracks must have at least one hit in the two innermost layers of the ITS and must have crossed a minimum of 70 out of total 159 rows along the transverse readout plane of the TPC. The maximum $\chi^{2}$ per space point in the TPC and ITS obtained from the track fit are required to be 4 and 36, respectively. To minimize the contribution of secondary charged particles, the distance of closest approach in the transverse plane of reconstructed tracks to the primary vertex (DCA$_{\mathrm{xy}}$) is required to be smaller than 7$\sigma$, where $\sigma$ is the DCA$_{\mathrm{xy}}$ resolution. The DCA$_{\mathrm{xy}}$ resolution is found to be \pt dependent and is parameterized as $\sigma = 0.0105+0.0350/(\pt)^{1.1}$. The DCA in the longitudinal direction is required to be smaller than 2 cm. Selected charged particles are further identified via the TPC and the Time Of Flight (TOF)~\cite{ALICE:2000xcm} detectors using their specific ionization energy loss  d$E/$d$x$ in the TPC and flight time measured in the TOF. Pions ($\pi$) and kaons (K) are identified with the condition that their specific energy loss lies within 2 standard deviations ($\sigma_{\mathrm{TPC}}$) (for $p$ $>$ 0.4), 4$\sigma_{\mathrm{TPC}}$ (for 0.3 $<$ $p$ $<$ 0.4) and 6$\sigma_{\mathrm{TPC}}$ (for $p$ $<$ 0.3) from their expected d$E$/d$x$, where $\sigma_{\mathrm{TPC}}$ corresponds to the d$E/$d$x$ resolution (typically $\approx$5\% of the measured d$E/$d$x$ value) of the TPC. Furthermore, if the hit for a track is available in the TOF, the measured time of flight is required to be within 3$\sigma$ from its expected value for each particle species~\cite{Carnesecchi:2018oss}. 

\section{Analysis details} 
\label{DA}
\begin{figure}[!hbt]
	\centering
	\begin{minipage}{.5\textwidth}
		\includegraphics[height=1.0\linewidth,width=1.0\linewidth]{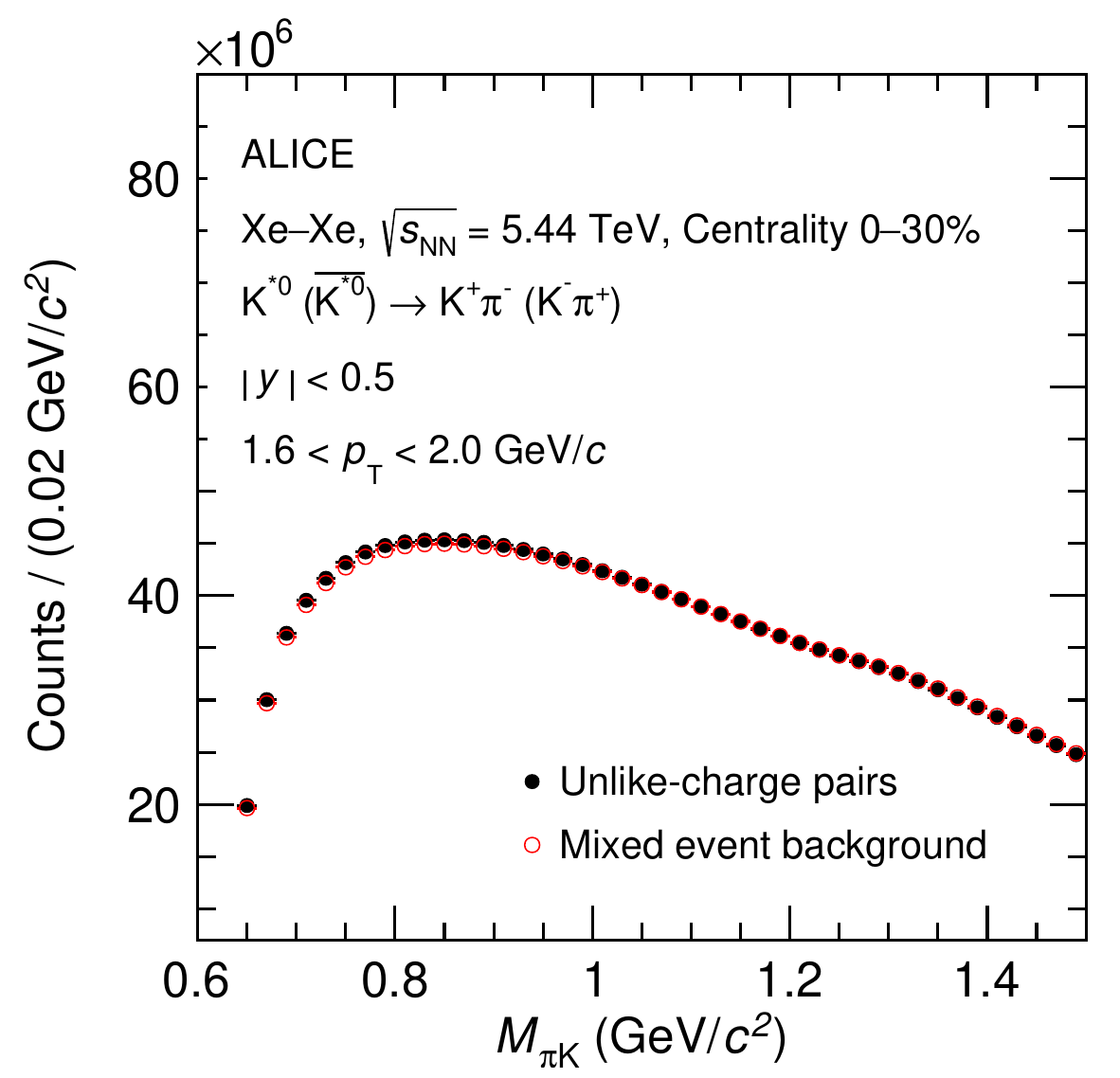}
	\end{minipage}%
	\begin{minipage}{.5\textwidth}
		\centering
		\includegraphics[height=1.0\linewidth,width=1.0\linewidth]{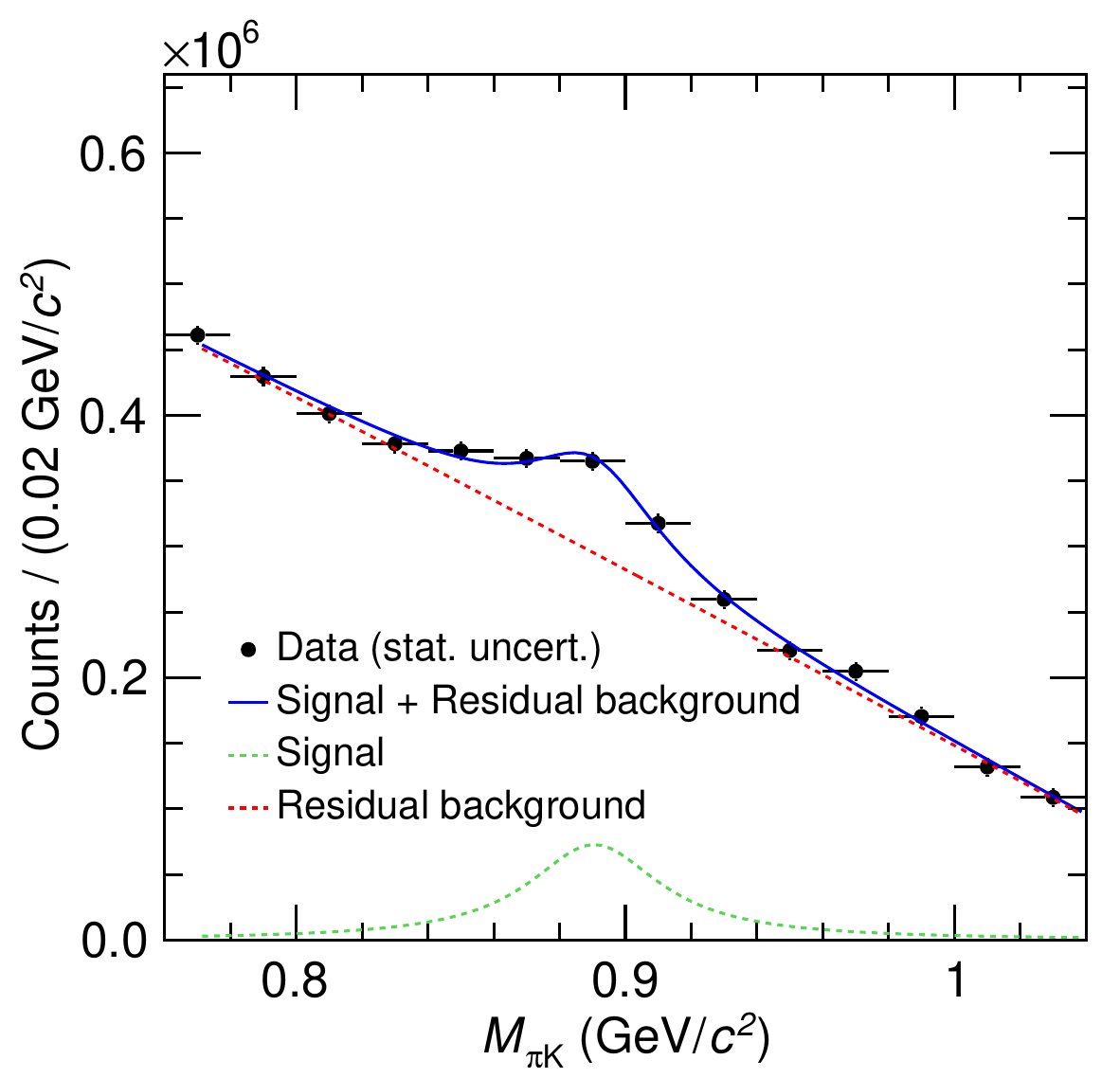}
	\end{minipage}	
	\caption{\label{fig:invmass} The left panel shows the invariant mass distribution of unlike sign $\pi$K pairs from the same and mixed events. The right panel shows the same but after the mixed-event background subtraction. The mixed event background subtracted invariant mass distribution is fitted with a combination of Breit--Wigner function~\cite{ALICE:2016sak} and second order polynomial distribution. The Breit--Wigner distribution represents the \Kstar signal and the second order polynomial describes the residual background.}
\end{figure}
The \NKSshort meson being a short-lived resonance is reconstructed using the invariant mass method~\cite{ALICE:2016sak} via its hadronic decay channel, K$^{*}(892)^{0}$ $\rightarrow$ K$^{\pm}\mathrm{\pi^{\mp}}$ with a branching ratio (BR) of 66$\%$~\cite{ParticleDataGroup:2022pth} for $|y| <$ 0.5. 

Oppositely charged kaons and pions are paired in the same event to reconstruct the resonance signal. The resulting invariant-mass distribution of unlike sign charge K\pion pair consists of a signal with significant combinatorial background, which is estimated using the mixed-event method~\cite{ALICE:2019etb} (for 0.4 $<$ \pt $<$ 0.8 GeV/$c$ in \XeXe collisions like-sign pairs from the same event~\cite{ALICE:2019etb} are used to get better description of the combinatorial background). The mixed-event invariant mass distribution is constructed by combining kaons from one event with oppositely charged pions from five other events. Only the events with a similar topology, such as an absolute difference in the $z$ coordinate of their collision vertex less than 1 cm and the centrality difference (for Xe$-$Xe) or multiplicity percentile (for pp) difference less than 5$\%$ are mixed. The mixed-event background is scaled to match the unlikesign foreground distribution in the invariant mass range 1.1--1.15 GeV/$c^{2}$. The left panel of Fig.~\ref{fig:invmass} shows the invariant-mass distribution of unlike charged $\pi$K pairs from the same-event along with the rescaled mixed-event background. The unlike sign $\pi$K invariant mass distribution with mixed-event background subtracted is shown in the right panel of Fig.~\ref{fig:invmass}. The combinatorial background subtracted invariant mass distribution consists of \Kstar signal and a residual background of correlated pairs. The correlated background pairs can originate from sources such as jets, decays of resonances with misidentified daughters, and decays with multiple daughters. The combinatorial background subtracted invariant mass distribution is fitted with a combination
of a non-relativistic Breit--Wigner distribution and a polynomial of second order. The Breit--Wigner distribution describes the signal of \Kstar, whereas the residual background is modelled using the polynomial function. The width of \NKSshort is kept fixed to its vacuum value in the fit procedure to estimate the signal, whereas it is allowed to vary freely to estimate the systematic uncertainty. Finally, raw yields of \Kstar in each \pt interval and event class is obtained from the integral of the Breit--Wigner distribution as done in Refs.~\cite{ALICE:2017ban, ALICE:2019xyr}.

The extracted raw yields ($N^{\mathrm{raw}}$) are further corrected for detector acceptance and reconstruction efficiency ($A\times\epsilon_{\mathrm{rec}}$) and BR of the decay channel. The $A\times\epsilon_{\mathrm{rec}}$ is estimated using dedicated Monte Carlo (MC) event generators, PYTHIA8~\cite{Skands:2014pea} for pp collisions and HIJING~\cite{Wang:1991hta} for Xe--Xe collisions, with particles propagated through a simulation of the ALICE detector using GEANT3~\cite{Brun:1994aa}. A weighting procedure of the $A\times\epsilon_{\mathrm{rec}}$ is further used to account for the variation of $A\times\epsilon_{\mathrm{rec}}$ over the width of a \pt interval in the measured spectrum and for the mismatch in the shape of the spectrum in data and MC simulation~\cite{ALICE:2021ptz}. The input \pt distribution in MC is adjusted to match the real
distribution using \pt-dependent weights. These are defined as the ratio between the measured \pt distribution after all corrections are applied and the default distribution in MC. In the first iteration, an appropriate fit function with parameters taken from similar analyses is used to parameterize the \pt shape. After all corrections, the \pt spectrum is fitted with the fit function again and the updated parameters are used to modify the weights in the next iteration. Such an iterative procedure is repeated until convergence. Finally, the yields are normalized by the number of accepted events ($N_{\mathrm{event}}^{\rm{acc}}$) to obtain the corrected \pt spectrum in different event classes. Measurements in pp collisions are further corrected for the event loss and the signal loss, evaluated from the MC simulation. The signal loss correction ($\mathrm{f_{SL}}$) for \Kstar is calculated for each multiplicity class by taking the ratio of the simulated \Kstar \pt spectrum before trigger and event selection with the corresponding \pt spectra after applying all the selections. The $\mathrm{f_{SL}}$ is dominant at low \pt in 70--100$\%$ multiplicity class with the maximum value of 22$\%$. The event loss correction ($\mathrm{f_{ev}}$) corresponds to the fraction of INEL $>$ 0 events that do not pass the event-selection criteria and is estimated in~\cite{ALICE:2019avo}. The $\mathrm{f_{ev}}$ is not particle and \pt dependent, and its value spans from 0.99 in 0--1$\%$ multiplicity class to 0.71 in 70--100$\%$ multiplicity class. The corrected \pt spectrum can be expressed as

\begin{equation}
	\frac{1}{N_{\mathrm{event}}} \frac{\mathrm{d}^{2}N}{\mathrm{d}y\mathrm{d}p_{\mathrm{T}}}= \frac{1}{N^{\rm{acc}}_{\rm{event}}}\frac{\mathrm{d}^{2}N^{\mathrm{raw}}}{\mathrm{d}y\mathrm{d}p_{\mathrm{T}}}\frac{\mathrm{f_{ev}f_{SL}}}{(\mathrm{A}\times\epsilon_{\mathrm{rec}})\mathrm{BR}}.
\end{equation} 

\section{Systematic uncertainties} 
\label{Section:Systematics}

Systematic uncertainties on the measured \Kstar yields originate from various sources, including the signal extraction method, track selection, and particle identification criteria, the method used to match track segments in the ITS with tracks in the TPC, as well as uncertainties in the material budget and interaction cross section. The resulting changes in the \Kstar yields for each \pt and multiplicity (centrality) interval, obtained from repeating the full analysis chain with the variations and corrections described below, are incorporated as systematic uncertainties. Table~\ref{tab:sysunc} provides a summary of the systematic uncertainties on the measured \Kstar yields. The reported uncertainties in the table are averaged over all centrality/multiplicity classes and presented for a low- and high-\pt interval.

\begin{table}[ht]
	\centering
	\caption{Systematic uncertainties on measured \Kstar yield in pp and Xe$-$Xe collisions at \ENpp and \ENxe respectively. The systematic uncertainties are shown for different sources for a low- and a high-\pt interval.}
	\begin{tabular}{|c|c|c|c|c|}	
		\hline
		Systematic variation & \multicolumn{2}{|c|}{pp [$p_{\mathrm{T}}$ (GeV/$c$)]} & \multicolumn{2}{|c|}{\XeXe [$p_{\mathrm{T}}$ (GeV/$c$)]}\\
		\hline
		& 0--0.4 & 10.0--14.0 & 0.4--0.8 &  8.0--12.0\\
		\hline
		Signal extraction ($\%$) &  7.4  & 9.6 & 12.7& 11.5\\
		\hline
		Primary track selection ($\%$) & 1.9 & 5.0& 7.2&7.1\\
		\hline
		Particle identification ($\%$) & 1.4  & 5.5& 7.1&7.8\\
		\hline
		ITS--TPC matching ($\%$) & 2& negl.& 6.4&8.6\\
		\hline
		Material budget ($\%$) & 1.8 & negl.& 1.4&negl.\\
		\hline
		Hadronic interaction ($\%$) & 2.6& negl. & 2.3 &negl.\\
		\hline	
		Total ($\%$) & 8.7& 12.3&17.6 & 17.8\\
		\hline	 
	\end{tabular}
	\label{tab:sysunc}	
\end{table}

To evaluate the signal extraction uncertainty, several factors are varied, such as fitting ranges, mixed-event background rescaling region, residual background fit functions, and yield extraction methods. The default case involved fixed-width fits to the invariant mass distributions, based on the background shape. To assess the systematic uncertainty, the boundaries of the fitting ranges are adjusted by 20 MeV/$c^{2}$ on both sides. The rescaling of the mixed-event background distribution is shifted to different ranges to examine its impact. The residual background is modeled using a third-order polynomial to study systematic effects. For the primary track selection, the criteria are varied following the procedure described in Ref.~\cite{ALICE:2019etb}. Uncertainties associated with the identification of primary daughter tracks are estimated by varying the selection criteria in the TPC and TOF. Furthermore, uncertainties related to the material budget and hadronic cross section are obtained from Ref.~\cite{ALICE:2019etb}. The total uncertainty, obtained by summing the uncertainties from each source in quadrature, is averaged over all multiplicity classes. In pp collisions, the total uncertainty ranges from 6.5\% to 12.3\%, while in \XeXe collisions, it ranges from 15\% to 18\%.

\section{Results} 
\label{Res}

\begin{figure}[!hbt]
	\centering
	\includegraphics[height=0.6\textwidth]{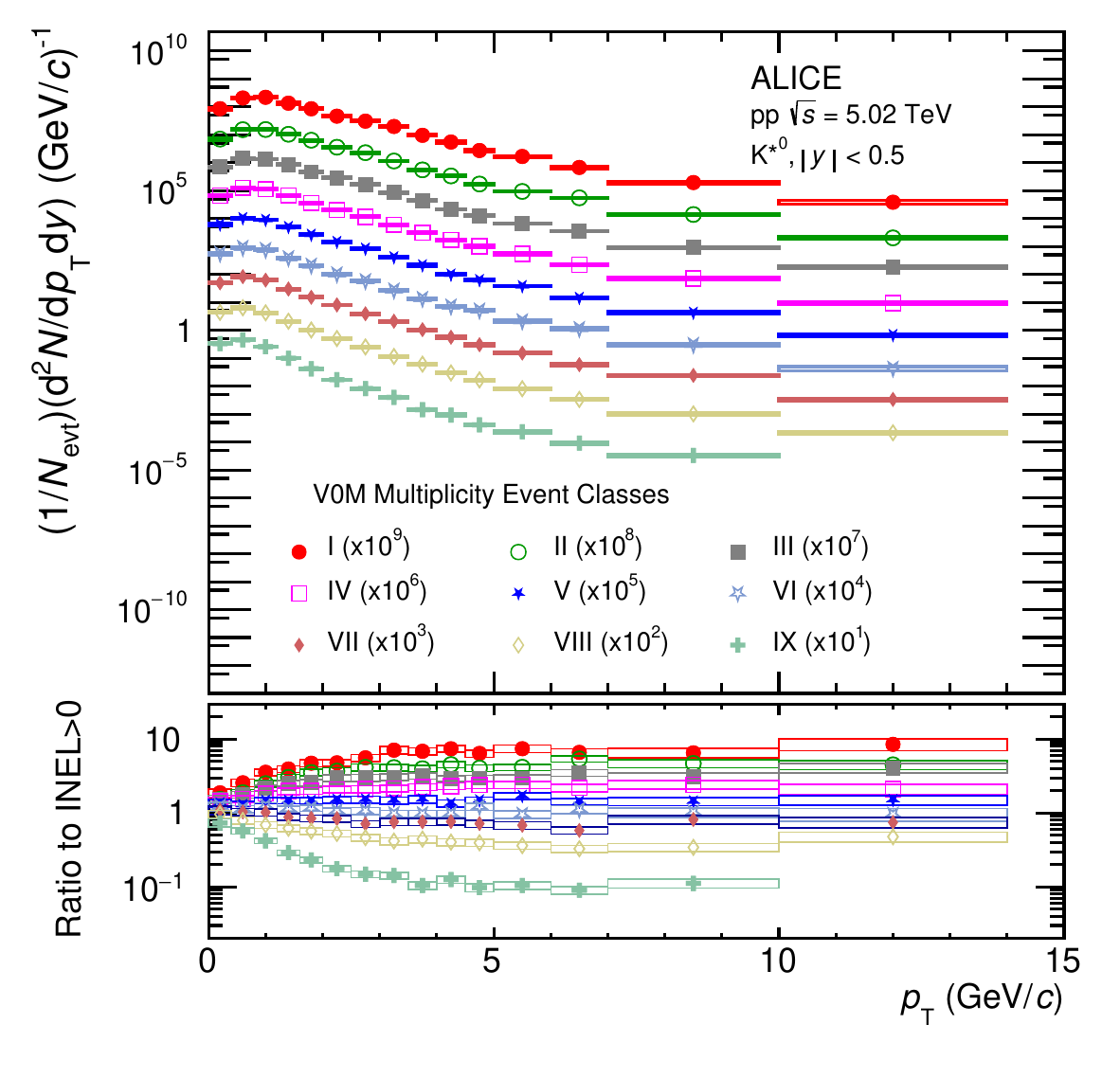}
	\caption{\label{fig:corrspectrapp} Upper panel: The $p_{\mathrm{T}}$ spectra of \Kstar in various multiplicity classes of pp collisions at \s $=$ 5.02 TeV. Lower panel: The ratios of the multiplicity-dependent \pt spectra to the multiplicity-integrated INEL$>$ 0 spectra. The statistical and systematic uncertainties are shown as bars and boxes, respectively.}
\end{figure}
\begin{figure}[!hbt]
	\centering
	\includegraphics[height=0.5\textwidth]{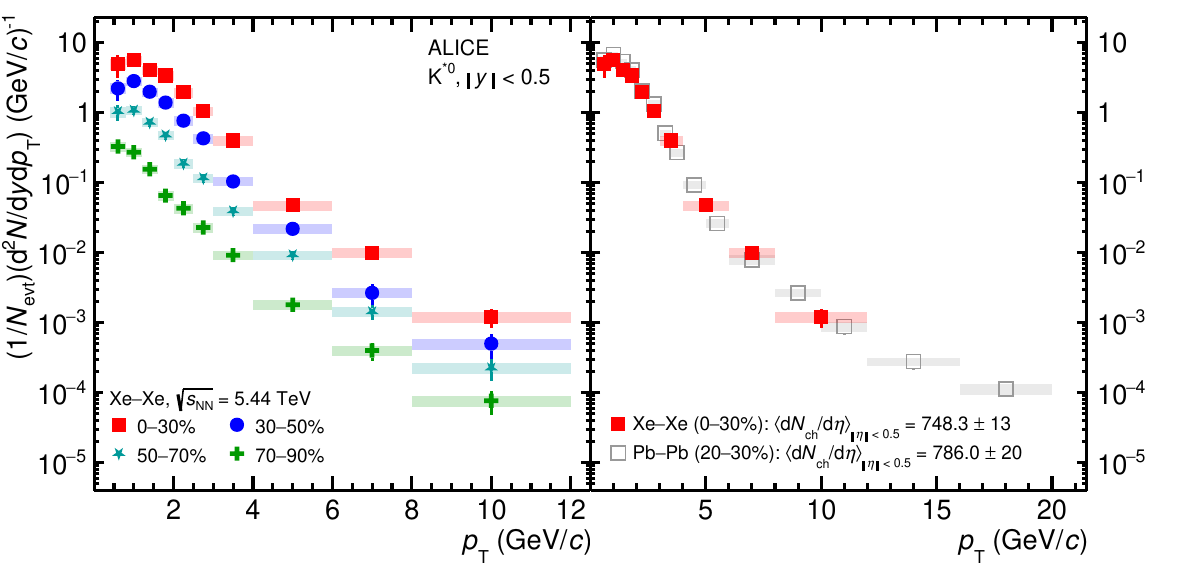}
	\caption{\label{fig:corrspectraXeXe} The left panel shows the \pt distributions of \Kstar meson in four different centrality classes of \XeXe collisions at \ENxe. The right panel shows the comparison between the \Kstar \pt spectrum in 0$-$30$\%$ \XeXe collisions at \snn $=$ 5.44 TeV and in 20$-$30$\%$ \PbPb~\cite{ALICE:2021ptz} collisions at \snn $=$ 5.02 TeV, both having similar multiplicities. The statistical and systematic uncertainties are shown by bars and boxes, respectively.}
\end{figure}

The \Kstar \pt spectra in pp collisions at \s $=$ 5.02 TeV for different multiplicity classes after all corrections mentioned in Sec.~\ref{DA} are shown in the upper panel of Fig.~\ref{fig:corrspectrapp}. The lower panel of Fig.~\ref{fig:corrspectrapp} shows the ratios of the \Kstar \pt spectra in different multiplicity classes to the corresponding spectrum in multiplicity integrated (INEL${>}$0) pp collisions. An increase in the inverse slopes of the \pt spectra from low to high multiplicity is clearly visible for \pt $<$ 4 GeV/$c$. However, at higher \pt, the spectra in different multiplicity classes have the same shape, indicating that the low \pt processes are primarily responsible for the change in the shape of the \pt spectra from low to high multiplicity classes. The corrected \pt distributions for \Kstar in four different centrality classes of \XeXe collisions at \snn $=$ 5.44 TeV are shown in the left panel of Fig.~\ref{fig:corrspectraXeXe}. The right panel of Fig.~\ref{fig:corrspectraXeXe} shows the comparison of the \Kstar \pt spectrum between \XeXe and \PbPb collisions with similar final-state charged-particle multiplicity. At similar multiplicity values, the \Kstar \pt distributions in \XeXe and \PbPb collisions are consistent within uncertainties. The final-state charged-particle multiplicity is a proxy of the volume of the produced matter. It is similar in the central collision of medium (Xe) and mid-central collisions of large (Pb) size nuclei. This indicates that the physics processes such as hadronic rescattering and radial flow, which determine the shape of the \pt distribution in heavy-ion collisions, have a similar effect on the \Kstar \pt spectra irrespective of the size of the colliding nuclei.

\begin{figure}[!hbt]
	\centering
	\includegraphics[height=0.5\textwidth]{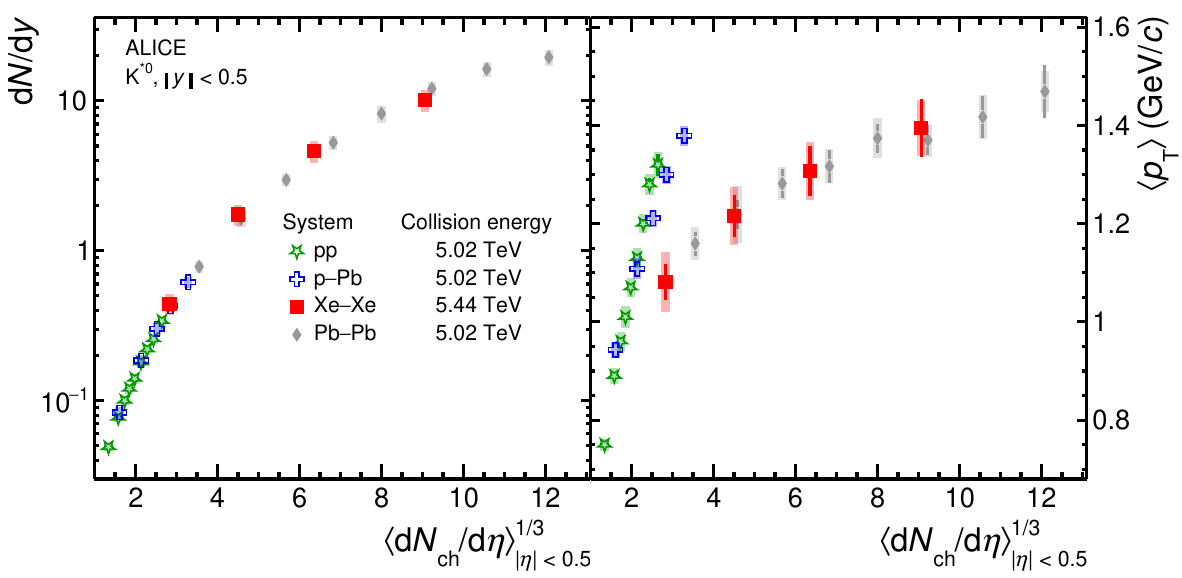}
	\caption{\label{fig:dNdYmpt} The \dndy (left panel) and \meanpt (right panel) of \Kstar as a function of \dndetaqube in pp collision at \s $=$ 5.02 TeV and in \XeXe collisions at \snn $=$ 5.44 TeV. Measurements are compared with the results obtained in p--Pb~\cite{ALICE:2016sak} and \PbPb~\cite{ALICE:2021ptz} collisions at \snn $=$ 5.02 TeV. Bars and shaded boxes correspond to the statistical and systematic uncertainties, respectively.}
\end{figure}
The transverse momentum integrated \Kstar yield \dndy and average transverse momentum \meanpt are extracted from the measured \pt spectrum and the extrapolation to the unmeasured regions using a blast-wave function~\cite{ALICE:2021ptz}. In pp collisions, \Kstar is measured down to \pt $=$ 0 GeV/$c$. Therefore, no low-\pt extrapolation is used to extract the \dndy and \meanpt in pp collisions. The contribution of the extrapolation on the extracted \dndy is $\approx$9\% ($\approx$13\%) in central (peripheral) \XeXe collisions. The systematic uncertainties on the extracted \dndy, and \meanpt are estimated by varying the data points randomly up and down within their systematic uncertainty to obtain the softest and hardest spectra. An additional systematic uncertainty due to the extrapolation of \pt spectra to \pt $=$ 0 GeV/$c$ is evaluated in Xe$-$Xe collisions by using different fit functions (Levy--Tsallis, Boltzmann) for the extrapolation~\cite{Tsallis:1987eu, ALICE:2020jsh}. The systematic uncertainty for the extrapolation is $\approx$2\% and $\approx$1.7\% on \dndy and \meanpt, respectively.

Figure~\ref{fig:dNdYmpt} shows the \dndy (left panel) and \meanpt (right panel) of \Kstar as a function of \dndetaqube in pp collisions at \s $=$ 5.02 TeV and in \XeXe collisions at \snn $=$ 5.44 TeV, where \dndetaqube is proportional to the linear (radial) path through the produced matter. Measurements are compared with the results obtained in p--Pb~\cite{ALICE:2016sak} and Pb--Pb collisions~\cite{ALICE:2021ptz} at \snn $=$ 5.02 TeV. A smooth evolution of the \dndy as a function of \dndetaqube is observed across all the collision systems. This suggests that \Kstar production is solely driven by final-state charged-particle multiplicity, which is used as a proxy for the system size~\cite{ALICE:2011dyt}. The \meanpt of \Kstar increases with \dndetaqube for all collision systems, indicating the increase of radial flow velocity from the low-multiplicity event class to the high-multiplicity event class. In contrast to the \dndy, intensive variable \meanpt shows a strong dependency on the colliding system and does not scale with charged-particle multiplicity across all collision systems. The \meanpt of \Kstar increases more steeply in small collision systems compared to heavy-ion collisions. For \dndetaqube $>$ 2 the following ordering of \meanpt is observed for a fixed multiplicity: \meanpt (pp) $>$ \meanpt (p--Pb) $>$ \meanpt (\XeXe) $\sim$ \meanpt (\PbPb). In the blast wave fit, where the fit parameters are interpreted in terms of a collective expansion, it is observed that small collision systems exhibit a larger pressure gradient and faster expansion of produced matter compared to heavy-ion collisions with similar charged-particle multiplicity~\cite{ALICE:2020nkc, Heinz:2019dbd}. Furthermore, the \meanpt of \Kstar in Xe$-$Xe and Pb$-$Pb collisions are comparable at similar \dndetaqube, suggesting similar dynamical evolution of the system produced in the collision of large and medium size nuclei at LHC energy.

\begin{figure}[!hbt]
	\centering
	\includegraphics[height=0.5\textwidth]{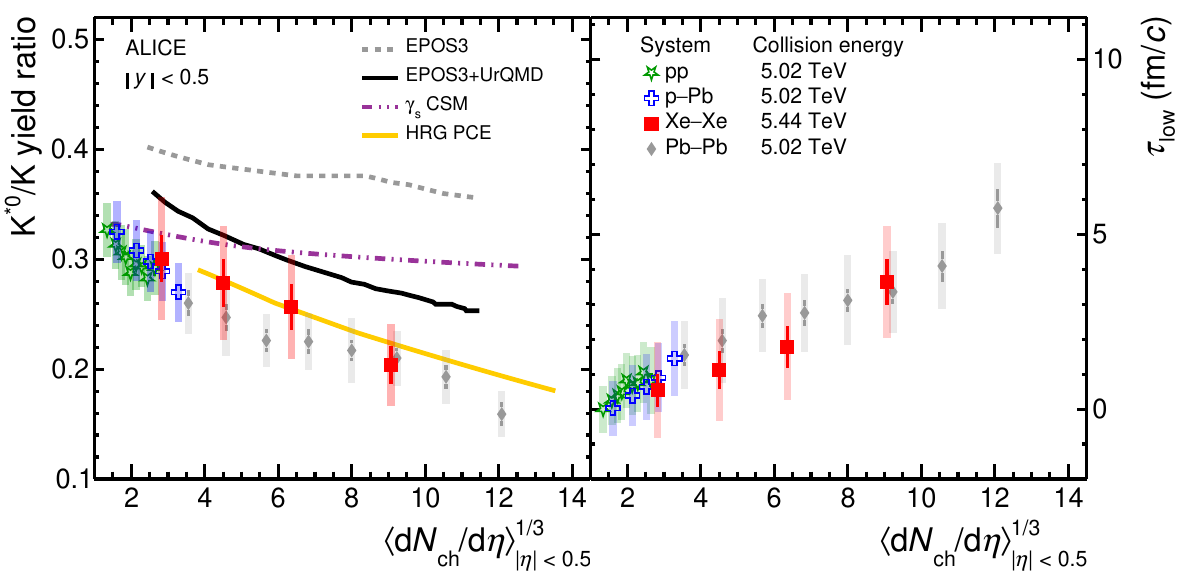}
	\caption{\label{fig:pr} The left panel shows the measured \Kstar/K yield ratio along with model calculation. The right panel shows the lower limit of hadronic phase lifetime as a function of \dndetaqube in different collision systems. Bars and shaded boxes represent the statistical and systematic uncertainties, respectively.}
\end{figure}
The left panel of Fig.~\ref{fig:pr} shows the \pt-integrated \Kstar/K yield ratio as a function of \dndetaqube. Measurements in \XeXe collisions are compared with the yield ratios obtained in pp, p--Pb~\cite{ALICE:2016sak} and Pb--Pb~\cite{ALICE:2021ptz} collisions at \snn $=$ 5.02 TeV. The kaon yields in pp collisions at \s $=$ 5.02 TeV are obtained through an extrapolation of kaon yields from pp collisions at \s $=$ 13 TeV~\cite{ALICE:2020nkc} and \s $=$ 7 TeV~\cite{ALICE:2018pal}. To perform this extrapolation, the yields at both \s $=$ 13 and \s $=$ 7 TeV are fitted as a function of \dndetaqube with a first-order polynomial. The resulting fit function value is then used to estimate the kaon yields at the corresponding \dndetaqube for \s $=$ 5.02 TeV. To assess the uncertainty in the yield estimation, a Gaussian distribution is constructed for each data point. The mean of the distribution corresponds to the value of the data point, while the standard deviation ($\sigma$) represents the associated statistical or systematic uncertainty. For each data point, a random value is sampled from its corresponding Gaussian distribution. It is assumed that the data points are uncorrelated with multiplicity. A linear fit is then applied to these randomly sampled values. This process is repeated thousands of times, generating multiple linear fits. The standard deviation of the fitting values obtained from these repetitions is considered as the uncertainty of the yield for a given multiplicity. The \Kstar/K yield ratio in different collision systems shows a smooth evolution with \dndetaqube, and is independent of the collision system at similar final-state charged-particle multiplicity. This further confirms the smooth evolution of hadron chemistry, observed for other light flavor hadrons~\cite{ALICE:2021lsv}. The \Kstar/K yield ratio decreases with increasing event multiplicity. This decrease in the \Kstar/K yield ratio can be understood as the rescattering of \Kstar meson's decay daughters inside the hadronic phase~\cite{ALICE:2019xyr}. Since the lifetime of \Kstar is comparable to that of the hadronic phase, its decay products scatter in their passage through the hadronic medium changing their momenta and hence affecting the reconstruction of the parent particle, thereby decreasing the measured yield. Measurements in heavy-ion collisions are further compared with the EPOS3 model calculations with and without the hadronic phase~\cite{Knospe:2015nva}. The EPOS3 model calculations are for \PbPb collisions at \snn $=$ 2.76 TeV, as no significant quantitative differences are expected between the two energies. In the presence of the hadronic phase, which is modelled by the UrQMD model~\cite{Bleicher:1999xi}, the EPOS3 generator qualitatively reproduces the multiplicity dependence of the \Kstar/K yield ratio. The canonical ensemble based statistical model ($\gamma_{\mathrm{s}}$ CSM)~\cite{Vovchenko:2019kes}, which successfully describes the production of other light flavor hadrons in small collision systems and heavy-ion collisions, does not explain the multiplicity dependence of \Kstar/K yield ratio. The yield ratio is suppressed compared to the $\gamma_{\mathrm{s}}$ CSM, and the suppression is more prominent in central \XeXe and \PbPb collisions. In the recent development of the Hadron Resonance Gas (HRG) model, the hadronic phase effect is modelled by a concept of partial chemical equilibrium (PCE)~\cite{Motornenko:2019jha}. In this model, decays and regenerations of the resonances obey the law of mass action, ensuring an equilibrium between the abundance of different resonances and their decay products. By applying the HRG-PCE calculation, the measured data points of particle ratios (\Kstar/K), in heavy-ion collisions can be accurately described. The \Kstar/K yield ratio can also be used to get an estimate of the lower bound of the hadronic phase lifetime $\tau$, i.e. the time between chemical and kinetic freeze-out. The \Kstar/K yield ratio at kinetic freeze-out can be expressed as
$[\mathrm{K}^{*0}/\mathrm{K}]_{\mathrm{kinetic}} = [\mathrm{K}^{*0}/\mathrm{K}]_{\mathrm{chemical}} \times e^{-\tau/\tau_{K^{*0}}}$, where $\tau_{K^{*0}}$ is the vacuum lifetime of \Kstar, taken to be 4.16 fm$/c$. 
The $[\mathrm{K}^{*0}/\mathrm{K}]$ yield ratio in the 70--100\% multiplicity class of pp collisions at \s $=$ 5.02 TeV is used as a proxy for the $[\mathrm{K}^{*0}/\mathrm{K}]_{\mathrm{chemical}}$ and the measured \Kstar/K yield ratio in different multiplicity or centrality classes of pp, p--Pb, \XeXe, and \PbPb collisions are used as $[\mathrm{K}^{*0}/\mathrm{K}]_{\mathrm{kinetic}}$. 
The above procedure estimates the lower bound of the $\tau$ with the assumption that there is no regeneration of \Kstar in the hadronic medium. The hadronic phase lifetime obtained with this simple model is further scaled by a Lorentz factor $\sqrt{1+(\frac{\langle p_{\mathrm{T}}\rangle}{\mathrm{mass~of~}\mathrm{K}^{*0}})^{2}}$ and the extracted $\tau$ values are shown in the right panel of Fig.~\ref{fig:pr} as a function of \dndetaqube. The hadronic phase lifetime evolves smoothly with multiplicity. The lifetime of the hadronic phases produced in \XeXe and \PbPb collisions are consistent with each other at similar charged-particle multiplicity.

\begin{figure}[!hbt]
	\centering
	\includegraphics[height=0.6\textwidth]{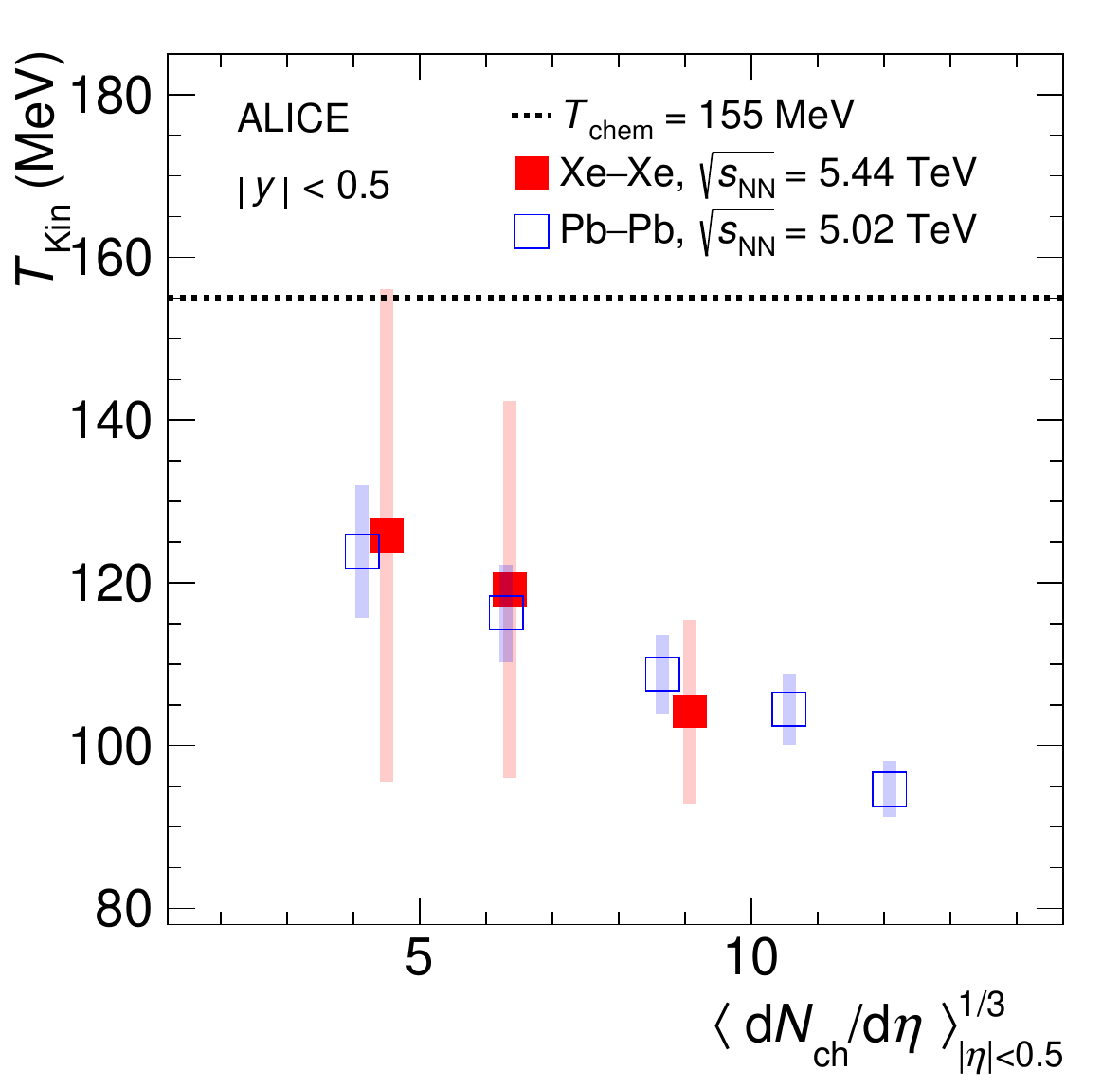}
	\caption{\label{fig:kintemp} The kinetic freeze-out temperature estimated using the fit of HRG-PCE model to the measured yields of $\pi^{\pm}$, K$^{\pm}$, p($\overline{\mathrm{p}}$), $\phi$, \Kstar in different centrality classes of Xe$-$Xe collisions at \ENxe. Results are compared with extracted kinetic freeze-out temperature in Pb$-$Pb collisions at \ENpbpb~\cite{ALICE:prottay}.}
\end{figure}
The time span by the hadronic phase is reflected in the temperature difference between the chemical and the kinetic freeze-out. The kinetic freeze-out temperature is extracted using the HRG-PCE~\cite{Motornenko:2019jha} model fit to the experimentally measured yields of $\pi^{\pm}$, K$^{\pm}$, p($\overline{\mathrm{p}}$), $\phi$~\cite{ALICE:2021ptz}, \Kstar in 0--30$\%$, 30--50$\%$ and 50--70$\%$ centrality classes for Xe--Xe collisions at \ENxe. The parameters of the fit are the baryon chemical potential, chemical freeze-out temperature, kinetic freeze-out temperature, and freeze-out volume of the system. The baryon chemical potential and chemical freeze-out temperature are fixed at 0 and 155 MeV, respectively, at LHC energies~\cite{Flor:2021olm, Andronic:2017pug, Vovchenko:2018fiy, Tawfik:2010kz}. Figure~\ref{fig:kintemp} shows the kinetic freeze-out temperature obtained from the HRG-PCE fit in Xe--Xe collisions, and the results are compared with the Pb--Pb measurements~\cite{ALICE:prottay}. The freeze-out temperature is found to increase systematically while moving from central to peripheral centrality class both for Xe--Xe and Pb--Pb collisions due to longer duration of hadronic phase, though the uncertainties are larger in \XeXe collisions. The freeze-out temperatures in both collision systems are consistent within uncertainties at similar charged-particle multiplicity. The difference between chemical and kinetic freeze-out temperature supports the presence of a hadronic phase with a finite lifetime in \XeXe collisions, a long-lived one in central collisions, and a short-lived one in peripheral collisions.

\begin{figure}[!hbt]
	\centering
	\includegraphics[height=0.5\textwidth]{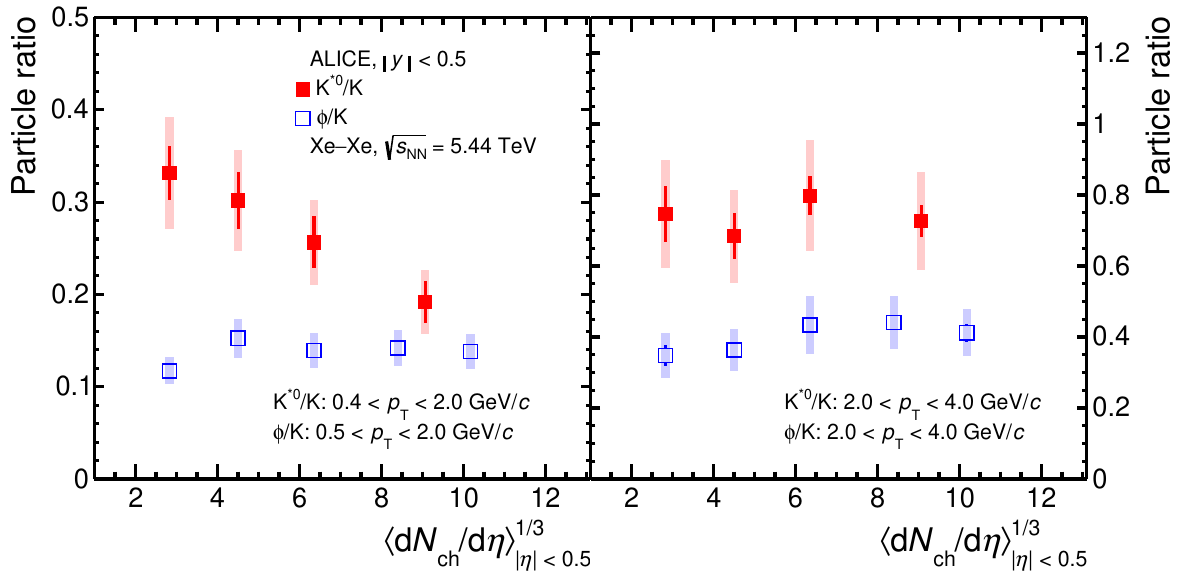}
	\caption{\label{fig:prdiff} The \Kstar/K and $\phi/$K yield ratios as a function of \dndetaqube in \XeXe collisions at \ENxe. The left and right panels show the measurements for a low-\pt and a high-\pt interval, respectively. Statistical and systematic uncertainties are represented by bars and shaded boxes.}
\end{figure}
Furthermore, to understand the \pt dependence of the hadronic rescattering effect, the \Kstar/K yield ratios in \XeXe collisions at \ENxe are shown in Fig.~\ref{fig:prdiff} for two different \pt intervals, 0.4 $ < p_{\rm{T}}<$ 2.0 GeV$/c$ and 2.0 $ < p_{\rm{T}}< $ 4.0 GeV$/c$. The results are also compared with the $\phi/$K~\cite{ALICE:2021lsv} yield ratio. In the low \pt range, the \Kstar/K yield ratio decreases from peripheral \XeXe collisions to central \XeXe collisions, whereas $\phi$/K remains more or less constant with system size. The observed low \pt suppression of measured \Kstar yield can be attributed to the rescattering effect of the decay products of \Kstar in the hadronic phase. The lifetime of $\phi$ mesons is one order of magnitude larger than that of \Kstar; therefore, the $\phi$ meson decay daughters are not expected to be affected by the rescattering in the hadronic phase. As a result, the $\phi/$K yield ratio remains constant within uncertainties across the whole range of multiplicities. In contrast to low \pt, at high \pt, both the \Kstar/K and $\phi$/K yield ratios remain flat as a function of \dndetaqube. This suggests that the rescattering effect is a low transverse momentum phenomenon.

\begin{figure}[!hbt]
	\centering
	\includegraphics[height=0.48\textwidth]{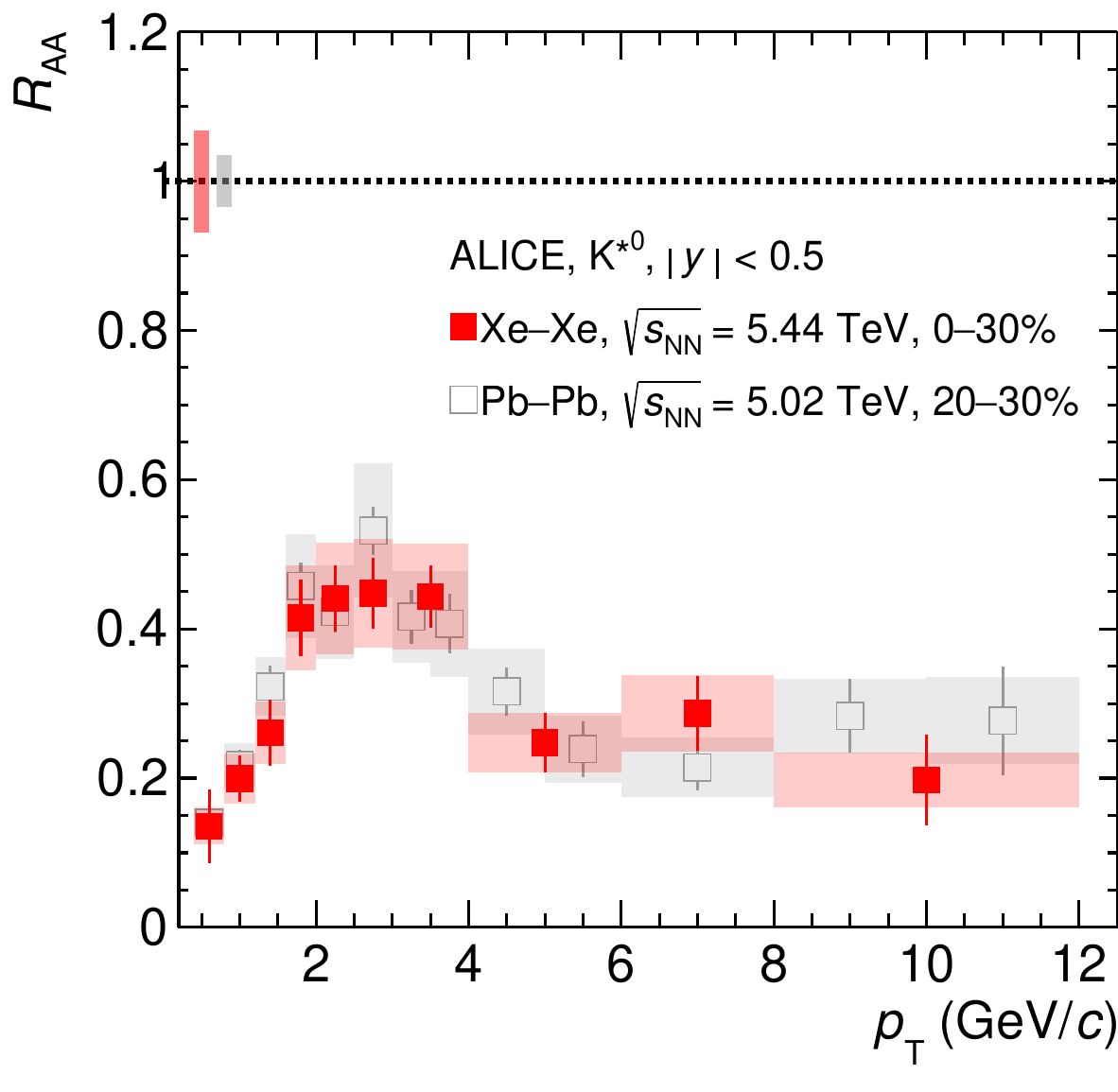}
	\includegraphics[height=0.48\textwidth]{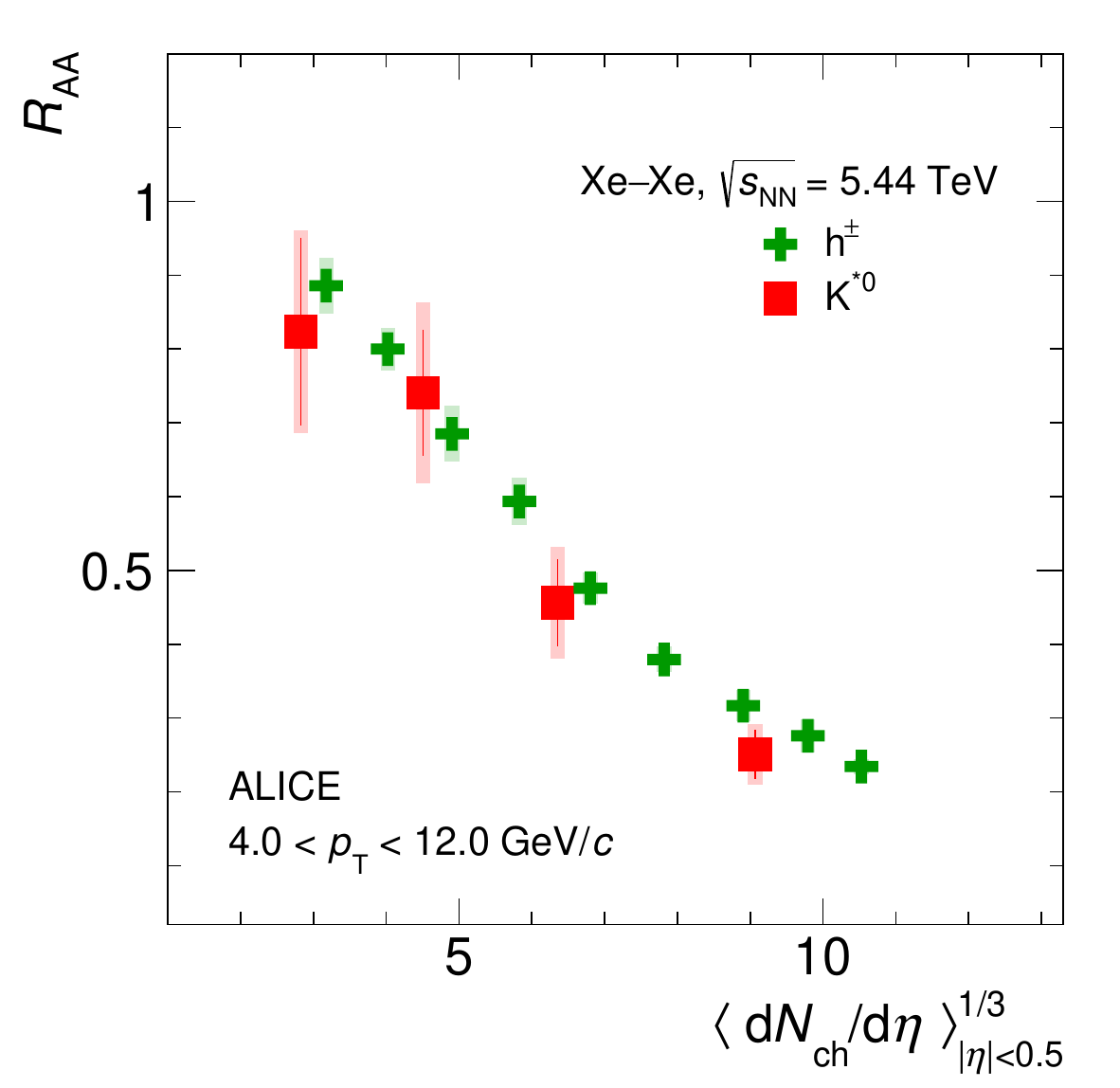}
	\caption{\label{fig:RAA1} The left panel shows the nuclear modification factor as a function of \pt for the \Kstar meson in 0$-$30$\%$ Xe$-$Xe collisions at \ENxe and in 20$-$30$\%$ Pb$-$Pb collisions~\cite{ALICE:2021ptz} at \ENpbpb. The right panel shows the R$_{\mathrm{AA}}$ of \Kstar as a function of \dndetaqube for 4.0 $<$ \pt $<$ 12.0 GeV/$c$ in \XeXe collisions. The results are compared to the $R_{\mathrm{AA}}$ of charged hadron~\cite{ALICE:2018hza}. Statistical and systematic uncertainties are represented by bars and shaded boxes.}
\end{figure}

The left panel of Fig.~\ref{fig:RAA1} shows the comparison of the nuclear modification factor $R_{\mathrm{AA}}$ of \Kstar in \XeXe and \PbPb systems at similar final-state charged-particle multiplicity. The $R_{\mathrm{AA}}$ values are found to be less than unity at high \pt in both systems. Similar $R_{\mathrm{AA}}$ is observed at both low momentum (hydro-like expansion) and high momentum (partonic energy loss) in Xe$-$Xe and Pb$-$Pb collisions at similar charged-particle multiplicity. The centrality dependence of energy loss is studied by measuring the \pt-integrated $R_{\mathrm{AA}}$ in the range 4.0 $<$ \pt $<$ 12.0 GeV$/c$. The \pt-integrated $R_{\mathrm{AA}}$ of \Kstar as a function of \dndetaqube is shown in the right panel of Fig.~\ref{fig:RAA1}. Measurements are compared with the results of charged hadrons and are found to be consistent within uncertainties. This suggests that jet quenching does not significantly affect the light-flavor particle species composition for the leading particles. The $R_{\mathrm{AA}}$ is smaller in central \XeXe collisions compared to the peripheral collisions. This reflects more energy loss via multiple partonic interactions in central collisions, as expected from the longer path length traversed by the hard partons in central collisions.

\section{Conclusion}
\label{Section:Summary}

The ALICE Collaboration has reported measurements of \Kstar meson at midrapidity ($|y|<$ 0.5) for different centrality and multiplicity classes in \XeXe and pp collisions at \ENxe and \ENpp, respectively. Both \pt-integrated \Kstar yield and \Kstar/K yield ratio are found to smoothly evolve with \dndetaqube, independent of the size of the colliding nuclei, confirming a universal scaling of hadron chemistry or relative abundance of hadron species with final-state charged-particle multiplicity at LHC energies. In contrast, the \meanpt, which depends on the radial expansion velocity of the produced matter, rises more steeply in smaller collision systems compared to the heavy-ion collisions. This indicates that the matter produced in small collision systems expands more rapidly compared to the system produced in heavy-ion collisions. 
The \Kstar/K ratio decreases with increasing final-state charged-particle multiplicity. This decrease in the \Kstar/K yield ratio can be attributed to the rescattering of decay daughters of \Kstar in the hadronic phase. In addition, the \pt-differential yield ratio \Kstar/K confirms the dominance of rescattering effect at low \pt. 
Moreover, the nuclear modification factor for \Kstar is similar in \XeXe and \PbPb collisions at similar charged-particle multiplicity indicating a scaling of the parton energy loss with final-state charged-particle multiplicity, independent of the size of the collision system.\\

The decreasing \Kstar/K ratio is qualitatively described by the EPOS3 model in presence of hadronic afterburner. The best description of the measurement is provided by the PCE based thermal model, which models the rescattering and the regeneration effect using the law of mass action. In contrast, the canonical ensemble based thermal model does not describe the measured \Kstar/K yield ratio.
Furthermore, the lower limit of hadronic phase lifetime is extracted using \Kstar/K yield ratios in different colliding systems. A smooth evolution of the lifetime is observed as a function of multiplicity. The kinetic freeze-out temperature is extracted using the HRG-PCE model. A higher temperature is obtained for more peripheral collisions implying an early decoupling of the produced hadrons.

\newenvironment{acknowledgement}{\relax}{\relax}
\begin{acknowledgement}
\section*{Acknowledgements}

The ALICE Collaboration would like to thank all its engineers and technicians for their invaluable contributions to the construction of the experiment and the CERN accelerator teams for the outstanding performance of the LHC complex.
The ALICE Collaboration gratefully acknowledges the resources and support provided by all Grid centres and the Worldwide LHC Computing Grid (WLCG) collaboration.
The ALICE Collaboration acknowledges the following funding agencies for their support in building and running the ALICE detector:
A. I. Alikhanyan National Science Laboratory (Yerevan Physics Institute) Foundation (ANSL), State Committee of Science and World Federation of Scientists (WFS), Armenia;
Austrian Academy of Sciences, Austrian Science Fund (FWF): [M 2467-N36] and Nationalstiftung f\"{u}r Forschung, Technologie und Entwicklung, Austria;
Ministry of Communications and High Technologies, National Nuclear Research Center, Azerbaijan;
Conselho Nacional de Desenvolvimento Cient\'{\i}fico e Tecnol\'{o}gico (CNPq), Financiadora de Estudos e Projetos (Finep), Funda\c{c}\~{a}o de Amparo \`{a} Pesquisa do Estado de S\~{a}o Paulo (FAPESP) and Universidade Federal do Rio Grande do Sul (UFRGS), Brazil;
Bulgarian Ministry of Education and Science, within the National Roadmap for Research Infrastructures 2020-2027 (object CERN), Bulgaria;
Ministry of Education of China (MOEC) , Ministry of Science \& Technology of China (MSTC) and National Natural Science Foundation of China (NSFC), China;
Ministry of Science and Education and Croatian Science Foundation, Croatia;
Centro de Aplicaciones Tecnol\'{o}gicas y Desarrollo Nuclear (CEADEN), Cubaenerg\'{\i}a, Cuba;
Ministry of Education, Youth and Sports of the Czech Republic, Czech Republic;
The Danish Council for Independent Research | Natural Sciences, the VILLUM FONDEN and Danish National Research Foundation (DNRF), Denmark;
Helsinki Institute of Physics (HIP), Finland;
Commissariat \`{a} l'Energie Atomique (CEA) and Institut National de Physique Nucl\'{e}aire et de Physique des Particules (IN2P3) and Centre National de la Recherche Scientifique (CNRS), France;
Bundesministerium f\"{u}r Bildung und Forschung (BMBF) and GSI Helmholtzzentrum f\"{u}r Schwerionenforschung GmbH, Germany;
General Secretariat for Research and Technology, Ministry of Education, Research and Religions, Greece;
National Research, Development and Innovation Office, Hungary;
Department of Atomic Energy Government of India (DAE), Department of Science and Technology, Government of India (DST), University Grants Commission, Government of India (UGC) and Council of Scientific and Industrial Research (CSIR), India;
National Research and Innovation Agency - BRIN, Indonesia;
Istituto Nazionale di Fisica Nucleare (INFN), Italy;
Japanese Ministry of Education, Culture, Sports, Science and Technology (MEXT) and Japan Society for the Promotion of Science (JSPS) KAKENHI, Japan;
Consejo Nacional de Ciencia (CONACYT) y Tecnolog\'{i}a, through Fondo de Cooperaci\'{o}n Internacional en Ciencia y Tecnolog\'{i}a (FONCICYT) and Direcci\'{o}n General de Asuntos del Personal Academico (DGAPA), Mexico;
Nederlandse Organisatie voor Wetenschappelijk Onderzoek (NWO), Netherlands;
The Research Council of Norway, Norway;
Commission on Science and Technology for Sustainable Development in the South (COMSATS), Pakistan;
Pontificia Universidad Cat\'{o}lica del Per\'{u}, Peru;
Ministry of Education and Science, National Science Centre and WUT ID-UB, Poland;
Korea Institute of Science and Technology Information and National Research Foundation of Korea (NRF), Republic of Korea;
Ministry of Education and Scientific Research, Institute of Atomic Physics, Ministry of Research and Innovation and Institute of Atomic Physics and University Politehnica of Bucharest, Romania;
Ministry of Education, Science, Research and Sport of the Slovak Republic, Slovakia;
National Research Foundation of South Africa, South Africa;
Swedish Research Council (VR) and Knut \& Alice Wallenberg Foundation (KAW), Sweden;
European Organization for Nuclear Research, Switzerland;
Suranaree University of Technology (SUT), National Science and Technology Development Agency (NSTDA), Thailand Science Research and Innovation (TSRI) and National Science, Research and Innovation Fund (NSRF), Thailand;
Turkish Energy, Nuclear and Mineral Research Agency (TENMAK), Turkey;
National Academy of  Sciences of Ukraine, Ukraine;
Science and Technology Facilities Council (STFC), United Kingdom;
National Science Foundation of the United States of America (NSF) and United States Department of Energy, Office of Nuclear Physics (DOE NP), United States of America.
In addition, individual groups or members have received support from:
European Research Council, Strong 2020 - Horizon 2020 (grant nos. 950692, 824093), European Union;
Academy of Finland (Center of Excellence in Quark Matter) (grant nos. 346327, 346328), Finland.

\end{acknowledgement}

\bibliographystyle{utphys}   
\bibliography{bibliography}

\newpage
\appendix

%
%

\section{The ALICE Collaboration}
\label{app:collab}
\begin{flushleft} 
\small

S.~Acharya\,\orcidlink{0000-0002-9213-5329}\,$^{\rm 128}$, 
D.~Adamov\'{a}\,\orcidlink{0000-0002-0504-7428}\,$^{\rm 87}$, 
G.~Aglieri Rinella\,\orcidlink{0000-0002-9611-3696}\,$^{\rm 33}$, 
M.~Agnello\,\orcidlink{0000-0002-0760-5075}\,$^{\rm 30}$, 
N.~Agrawal\,\orcidlink{0000-0003-0348-9836}\,$^{\rm 52}$, 
Z.~Ahammed\,\orcidlink{0000-0001-5241-7412}\,$^{\rm 136}$, 
S.~Ahmad\,\orcidlink{0000-0003-0497-5705}\,$^{\rm 16}$, 
S.U.~Ahn\,\orcidlink{0000-0001-8847-489X}\,$^{\rm 72}$, 
I.~Ahuja\,\orcidlink{0000-0002-4417-1392}\,$^{\rm 38}$, 
A.~Akindinov\,\orcidlink{0000-0002-7388-3022}\,$^{\rm 142}$
% ...

\section*{Affiliation Notes}

$^{\rm I}$ Deceased\\
$^{\rm II}$ Also at: Max-Planck-Institut fur Physik, Munich, Germany\\
$^{\rm III}$ Also at: Italian National Agency for New Technologies, Energy and Sustainable Economic Development (ENEA), Bologna, Italy\\
$^{\rm IV}$ Also at: Dipartimento DET del Politecnico di Torino, Turin, Italy\\
$^{\rm V}$ Also at: Department of Applied Physics, Aligarh Muslim University, Aligarh, India\\
$^{\rm VI}$ Also at: Institute of Theoretical Physics, University of Wroclaw, Poland\\
$^{\rm VII}$ Also at: An institution covered by a cooperation agreement with CERN\\

\section*{Collaboration Institutes}

$^{1}$ A.I. Alikhanyan National Science Laboratory (Yerevan Physics Institute) Foundation, Yerevan, Armenia\\
$^{2}$ AGH University of Krakow, Cracow, Poland\\
$^{3}$ Bogolyubov Institute for Theoretical Physics, National Academy of Sciences of Ukraine, Kiev, Ukraine\\
$^{4}$ Bose Institute, Department of Physics  and Centre for Astroparticle Physics and Space Science (CAPSS), Kolkata, India\\
$^{5}$ California Polytechnic State University, San Luis Obispo, California, United States\\
$^{6}$ Central China Normal University, Wuhan, China\\
$^{7}$ Centro de Aplicaciones Tecnol\'{o}gicas y Desarrollo Nuclear (CEADEN), Havana, Cuba\\
$^{8}$ Centro de Investigaci\'{o}n y de Estudios Avanzados (CINVESTAV), Mexico City and M\'{e}rida, Mexico\\
$^{9}$ Chicago State University, Chicago, Illinois, United States\\
$^{10}$ China Institute of Atomic Energy, Beijing, China\\
$^{11}$ China University of Geosciences, Wuhan, China\\
$^{12}$ Chungbuk National University, Cheongju, Republic of Korea\\
$^{13}$ Comenius University Bratislava, Faculty of Mathematics, Physics and Informatics, Bratislava, Slovak Republic\\
$^{14}$ COMSATS University Islamabad, Islamabad, Pakistan\\
$^{15}$ Creighton University, Omaha, Nebraska, United States\\
$^{16}$ Department of Physics, Aligarh Muslim University, Aligarh, India\\
$^{17}$ Department of Physics, Pusan National University, Pusan, Republic of Korea\\
$^{18}$ Department of Physics, Sejong University, Seoul, Republic of Korea\\
$^{19}$ Department of Physics, University of California, Berkeley, California, United States\\
$^{20}$ Department of Physics, University of Oslo, Oslo, Norway\\
$^{21}$ Department of Physics and Technology, University of Bergen, Bergen, Norway\\
$^{22}$ Dipartimento di Fisica, Universit\`{a} di Pavia, Pavia, Italy\\
$^{23}$ Dipartimento di Fisica dell'Universit\`{a} and Sezione INFN, Cagliari, Italy\\
$^{24}$ Dipartimento di Fisica dell'Universit\`{a} and Sezione INFN, Trieste, Italy\\
$^{25}$ Dipartimento di Fisica dell'Universit\`{a} and Sezione INFN, Turin, Italy\\
$^{26}$ Dipartimento di Fisica e Astronomia dell'Universit\`{a} and Sezione INFN, Bologna, Italy\\
$^{27}$ Dipartimento di Fisica e Astronomia dell'Universit\`{a} and Sezione INFN, Catania, Italy\\
$^{28}$ Dipartimento di Fisica e Astronomia dell'Universit\`{a} and Sezione INFN, Padova, Italy\\
$^{29}$ Dipartimento di Fisica `E.R.~Caianiello' dell'Universit\`{a} and Gruppo Collegato INFN, Salerno, Italy\\
$^{30}$ Dipartimento DISAT del Politecnico and Sezione INFN, Turin, Italy\\
$^{31}$ Dipartimento di Scienze MIFT, Universit\`{a} di Messina, Messina, Italy\\
$^{32}$ Dipartimento Interateneo di Fisica `M.~Merlin' and Sezione INFN, Bari, Italy\\
$^{33}$ European Organization for Nuclear Research (CERN), Geneva, Switzerland\\
$^{34}$ Faculty of Electrical Engineering, Mechanical Engineering and Naval Architecture, University of Split, Split, Croatia\\
$^{35}$ Faculty of Engineering and Science, Western Norway University of Applied Sciences, Bergen, Norway\\
$^{36}$ Faculty of Nuclear Sciences and Physical Engineering, Czech Technical University in Prague, Prague, Czech Republic\\
$^{37}$ Faculty of Physics, Sofia University, Sofia, Bulgaria\\
$^{38}$ Faculty of Science, P.J.~\v{S}af\'{a}rik University, Ko\v{s}ice, Slovak Republic\\
$^{39}$ Frankfurt Institute for Advanced Studies, Johann Wolfgang Goethe-Universit\"{a}t Frankfurt, Frankfurt, Germany\\
$^{40}$ Fudan University, Shanghai, China\\
$^{41}$ Gangneung-Wonju National University, Gangneung, Republic of Korea\\
$^{42}$ Gauhati University, Department of Physics, Guwahati, India\\
$^{43}$ Helmholtz-Institut f\"{u}r Strahlen- und Kernphysik, Rheinische Friedrich-Wilhelms-Universit\"{a}t Bonn, Bonn, Germany\\
$^{44}$ Helsinki Institute of Physics (HIP), Helsinki, Finland\\
$^{45}$ High Energy Physics Group,  Universidad Aut\'{o}noma de Puebla, Puebla, Mexico\\
$^{46}$ Horia Hulubei National Institute of Physics and Nuclear Engineering, Bucharest, Romania\\
$^{47}$ HUN-REN Wigner Research Centre for Physics, Budapest, Hungary\\
$^{48}$ Indian Institute of Technology Bombay (IIT), Mumbai, India\\
$^{49}$ Indian Institute of Technology Indore, Indore, India\\
$^{50}$ INFN, Laboratori Nazionali di Frascati, Frascati, Italy\\
$^{51}$ INFN, Sezione di Bari, Bari, Italy\\
$^{52}$ INFN, Sezione di Bologna, Bologna, Italy\\
$^{53}$ INFN, Sezione di Cagliari, Cagliari, Italy\\
$^{54}$ INFN, Sezione di Catania, Catania, Italy\\
$^{55}$ INFN, Sezione di Padova, Padova, Italy\\
$^{56}$ INFN, Sezione di Pavia, Pavia, Italy\\
$^{57}$ INFN, Sezione di Torino, Turin, Italy\\
$^{58}$ INFN, Sezione di Trieste, Trieste, Italy\\
$^{59}$ Inha University, Incheon, Republic of Korea\\
$^{60}$ Institute for Gravitational and Subatomic Physics (GRASP), Utrecht University/Nikhef, Utrecht, Netherlands\\
$^{61}$ Institute of Experimental Physics, Slovak Academy of Sciences, Ko\v{s}ice, Slovak Republic\\
$^{62}$ Institute of Physics, Homi Bhabha National Institute, Bhubaneswar, India\\
$^{63}$ Institute of Physics of the Czech Academy of Sciences, Prague, Czech Republic\\
$^{64}$ Institute of Space Science (ISS), Bucharest, Romania\\
$^{65}$ Institut f\"{u}r Kernphysik, Johann Wolfgang Goethe-Universit\"{a}t Frankfurt, Frankfurt, Germany\\
$^{66}$ Instituto de Ciencias Nucleares, Universidad Nacional Aut\'{o}noma de M\'{e}xico, Mexico City, Mexico\\
$^{67}$ Instituto de F\'{i}sica, Universidade Federal do Rio Grande do Sul (UFRGS), Porto Alegre, Brazil\\
$^{68}$ Instituto de F\'{\i}sica, Universidad Nacional Aut\'{o}noma de M\'{e}xico, Mexico City, Mexico\\
$^{69}$ iThemba LABS, National Research Foundation, Somerset West, South Africa\\
$^{70}$ Jeonbuk National University, Jeonju, Republic of Korea\\
$^{71}$ Johann-Wolfgang-Goethe Universit\"{a}t Frankfurt Institut f\"{u}r Informatik, Fachbereich Informatik und Mathematik, Frankfurt, Germany\\
$^{72}$ Korea Institute of Science and Technology Information, Daejeon, Republic of Korea\\
$^{73}$ KTO Karatay University, Konya, Turkey\\
$^{74}$ Laboratoire de Physique Subatomique et de Cosmologie, Universit\'{e} Grenoble-Alpes, CNRS-IN2P3, Grenoble, France\\
$^{75}$ Lawrence Berkeley National Laboratory, Berkeley, California, United States\\
$^{76}$ Lund University Department of Physics, Division of Particle Physics, Lund, Sweden\\
$^{77}$ Nagasaki Institute of Applied Science, Nagasaki, Japan\\
$^{78}$ Nara Women{'}s University (NWU), Nara, Japan\\
$^{79}$ National and Kapodistrian University of Athens, School of Science, Department of Physics , Athens, Greece\\
$^{80}$ National Centre for Nuclear Research, Warsaw, Poland\\
$^{81}$ National Institute of Science Education and Research, Homi Bhabha National Institute, Jatni, India\\
$^{82}$ National Nuclear Research Center, Baku, Azerbaijan\\
$^{83}$ National Research and Innovation Agency - BRIN, Jakarta, Indonesia\\
$^{84}$ Niels Bohr Institute, University of Copenhagen, Copenhagen, Denmark\\
$^{85}$ Nikhef, National institute for subatomic physics, Amsterdam, Netherlands\\
$^{86}$ Nuclear Physics Group, STFC Daresbury Laboratory, Daresbury, United Kingdom\\
$^{87}$ Nuclear Physics Institute of the Czech Academy of Sciences, Husinec-\v{R}e\v{z}, Czech Republic\\
$^{88}$ Oak Ridge National Laboratory, Oak Ridge, Tennessee, United States\\
$^{89}$ Ohio State University, Columbus, Ohio, United States\\
$^{90}$ Physics department, Faculty of science, University of Zagreb, Zagreb, Croatia\\
$^{91}$ Physics Department, Panjab University, Chandigarh, India\\
$^{92}$ Physics Department, University of Jammu, Jammu, India\\
$^{93}$ Physics Program and International Institute for Sustainability with Knotted Chiral Meta Matter (SKCM2), Hiroshima University, Hiroshima, Japan\\
$^{94}$ Physikalisches Institut, Eberhard-Karls-Universit\"{a}t T\"{u}bingen, T\"{u}bingen, Germany\\
$^{95}$ Physikalisches Institut, Ruprecht-Karls-Universit\"{a}t Heidelberg, Heidelberg, Germany\\
$^{96}$ Physik Department, Technische Universit\"{a}t M\"{u}nchen, Munich, Germany\\
$^{97}$ Politecnico di Bari and Sezione INFN, Bari, Italy\\
$^{98}$ Research Division and ExtreMe Matter Institute EMMI, GSI Helmholtzzentrum f\"ur Schwerionenforschung GmbH, Darmstadt, Germany\\
$^{99}$ Saga University, Saga, Japan\\
$^{100}$ Saha Institute of Nuclear Physics, Homi Bhabha National Institute, Kolkata, India\\
$^{101}$ School of Physics and Astronomy, University of Birmingham, Birmingham, United Kingdom\\
$^{102}$ Secci\'{o}n F\'{\i}sica, Departamento de Ciencias, Pontificia Universidad Cat\'{o}lica del Per\'{u}, Lima, Peru\\
$^{103}$ Stefan Meyer Institut f\"{u}r Subatomare Physik (SMI), Vienna, Austria\\
$^{104}$ SUBATECH, IMT Atlantique, Nantes Universit\'{e}, CNRS-IN2P3, Nantes, France\\
$^{105}$ Sungkyunkwan University, Suwon City, Republic of Korea\\
$^{106}$ Suranaree University of Technology, Nakhon Ratchasima, Thailand\\
$^{107}$ Technical University of Ko\v{s}ice, Ko\v{s}ice, Slovak Republic\\
$^{108}$ The Henryk Niewodniczanski Institute of Nuclear Physics, Polish Academy of Sciences, Cracow, Poland\\
$^{109}$ The University of Texas at Austin, Austin, Texas, United States\\
$^{110}$ Universidad Aut\'{o}noma de Sinaloa, Culiac\'{a}n, Mexico\\
$^{111}$ Universidade de S\~{a}o Paulo (USP), S\~{a}o Paulo, Brazil\\
$^{112}$ Universidade Estadual de Campinas (UNICAMP), Campinas, Brazil\\
$^{113}$ Universidade Federal do ABC, Santo Andre, Brazil\\
$^{114}$ Universitatea Nationala de Stiinta si Tehnologie Politehnica Bucuresti, Bucharest, Romania\\
$^{115}$ University of Cape Town, Cape Town, South Africa\\
$^{116}$ University of Derby, Derby, United Kingdom\\
$^{117}$ University of Houston, Houston, Texas, United States\\
$^{118}$ University of Jyv\"{a}skyl\"{a}, Jyv\"{a}skyl\"{a}, Finland\\
$^{119}$ University of Kansas, Lawrence, Kansas, United States\\
$^{120}$ University of Liverpool, Liverpool, United Kingdom\\
$^{121}$ University of Science and Technology of China, Hefei, China\\
$^{122}$ University of South-Eastern Norway, Kongsberg, Norway\\
$^{123}$ University of Tennessee, Knoxville, Tennessee, United States\\
$^{124}$ University of the Witwatersrand, Johannesburg, South Africa\\
$^{125}$ University of Tokyo, Tokyo, Japan\\
$^{126}$ University of Tsukuba, Tsukuba, Japan\\
$^{127}$ Universit\"{a}t M\"{u}nster, Institut f\"{u}r Kernphysik, M\"{u}nster, Germany\\
$^{128}$ Universit\'{e} Clermont Auvergne, CNRS/IN2P3, LPC, Clermont-Ferrand, France\\
$^{129}$ Universit\'{e} de Lyon, CNRS/IN2P3, Institut de Physique des 2 Infinis de Lyon, Lyon, France\\
$^{130}$ Universit\'{e} de Strasbourg, CNRS, IPHC UMR 7178, F-67000 Strasbourg, France, Strasbourg, France\\
$^{131}$ Universit\'{e} Paris-Saclay, Centre d'Etudes de Saclay (CEA), IRFU, D\'{e}partment de Physique Nucl\'{e}aire (DPhN), Saclay, France\\
$^{132}$ Universit\'{e}  Paris-Saclay, CNRS/IN2P3, IJCLab, Orsay, France\\
$^{133}$ Universit\`{a} degli Studi di Foggia, Foggia, Italy\\
$^{134}$ Universit\`{a} del Piemonte Orientale, Vercelli, Italy\\
$^{135}$ Universit\`{a} di Brescia, Brescia, Italy\\
$^{136}$ Variable Energy Cyclotron Centre, Homi Bhabha National Institute, Kolkata, India\\
$^{137}$ Warsaw University of Technology, Warsaw, Poland\\
$^{138}$ Wayne State University, Detroit, Michigan, United States\\
$^{139}$ Yale University, New Haven, Connecticut, United States\\
$^{140}$ Yonsei University, Seoul, Republic of Korea\\
$^{141}$  Zentrum  f\"{u}r Technologie und Transfer (ZTT), Worms, Germany\\
$^{142}$ Affiliated with an institute covered by a cooperation agreement with CERN\\
$^{143}$ Affiliated with an international laboratory covered by a cooperation agreement with CERN.\\

\end{flushleft} 

\end{document}